\documentclass[aps,prd,reprint,amsmath,amssymb,preprintnumbers,superscriptaddress,nofootinbib]{revtex4-1}
\usepackage[colorlinks]{hyperref}
\usepackage{graphicx}
\usepackage{here}

\begin{document}


\preprint{CTPU-PTC-18-14}
\title{
Machine learning for bounce calculation
}
\renewcommand{\thefootnote}{\alph{footnote}}

\author{
Ryusuke Jinno
}

\affiliation{
Center for Theoretical Physics of the Universe, Institute for Basic Science (IBS), Daejeon 34126, Korea
}

\begin{abstract}
We study the possibility of using machine learning for the calculation of the bounce action in quantum tunneling.
Adopting supervised learning, we train neural network to give the bounce action from a given potential.
It is found that, for one-dimensional tunneling, even a simple neural network performs at a percent level. 
\end{abstract}

\maketitle

\section{Introduction}

Machine learning is one of the most active fields in computer science.
It is producing remarkable advances in image and speech recognition, 
natural language processing, game playing, robot control and 
many other domains~\cite{LeCun:2015aa,Jordan:2015aa,Nielsen:2015aa}.
Its application to physics now ranges, for example, over 
collider physics~\cite{
Baldi:2014aa,
Baldi:2015aa,
Searcy:2016aa,
de-Oliveira:2016aa,
Komiske:2017aa,
Kasieczka:2017aa,
Erdmann:2017aa,
Barberio:2017aa,
Cohen:2018aa,
Demir:2018aa,
Guo:2018aa}, 
detection of phase transition or phase classification of matter~\cite{
Carrasquilla:2017aa,
Wang:2016aa,
Broecker:2017aa,
Chng:2017aa,
Deng:2017aa,
Tanaka:2017aa,
Ohtsuki:2016aa,
van-Nieuwenburg:2017aa,
Zhang:2017aa,
Pang:2016aa,
Ohtsuki:2017aa,
Wetzel:2017ab,
Hu:2017aa,
Schindler:2017aa,
Ponte:2017aa,
Wetzel:2017aa,
Mills:2018aa,
Costa:2017aa,
Mano:2017aa,
Liu:2017ab,
Venderley:2017aa,
Li:2017aa,
van-Nieuwenburg:2017ab},
gravitational-wave data analysis~\cite{
Biswas:2013aa,
Baker:2015aa,
Powell:2015aa,
Carrillo:2016aa,
Mukund:2017aa},
cosmic structure formation~\cite{
Kamdar:2016aa,
Kamdar:2016ab,
Agarwal:2018aa,
Nadler:2017aa,
Lucie-Smith:2018aa},
string landscape~\cite{
He:2017aa,
He:2017ab,
Ruehle:2017aa,
Carifio:2017aa,
Liu:2017aa,
Carifio:2017ab,
Wang:2018aa},
and
holography~\cite{Gan:2017aa,You:2018aa,Hashimoto:2018aa},
just to name a few.

In the present letter we apply machine learning to the calculation of the bounce action,
which appears in tunneling phenomena in quantum field theory. 
From phenomenological or cosmological viewpoint, 
such tunneling often accompanies first-order phase transition in the early Universe,
which is interesting because of its possible connection to the baryon asymmetry of the Universe~\cite{Kuzmin:1985aa}
and the production of gravitational waves~\cite{Witten:1984aa,Hogan:1986aa}.

In phenomenological studies on such first-order phase transitions,
we often have to calculate the tunneling rate of the scalar field $\phi$.
The tunneling rate is dominantly determined by a configuration called ``bounce,"
and the resulting tunneling rate is derived from an integration of this configuration called bounce action.
Since the traditional method to calculate this quantity can sometimes be numerically costly,
as explained in Sec.~\ref{sec:Image}, 
we investigate the possibility of using machine learning for the calculation of this quantity in the present letter.

The organization of the letter is as follows.
We first make our idea clear in Sec.~\ref{sec:Image}.
We next explain the dataset, setup of the neural network, and the obtained results in Sec.~\ref{sec:Analysis}.
Finally we summarize in Sec.~\ref{sec:DC}.

\section{Bounce calculation as image recognition}
\label{sec:Image}

\begin{figure}
\centerline{
\includegraphics[width=\columnwidth]{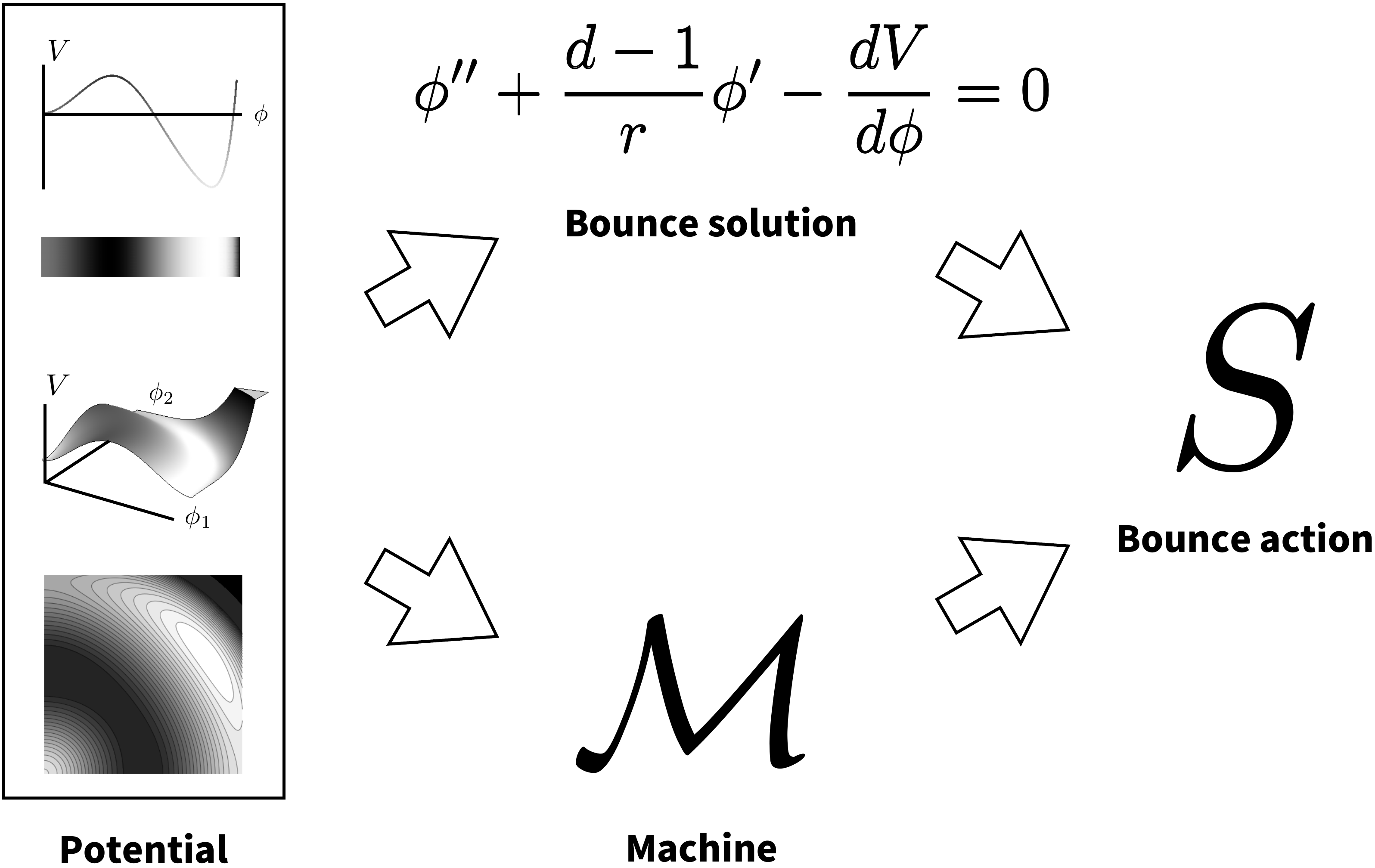}
}
\caption {\small
The main point in the present letter.
The standard procedure to calculate the bounce action is 
to first solve the bounce equation of motion and then integrate the obtained configuration (upper path).
However, a good machine may be able to learn the relation between the potential and the bounce action
without knowing the bounce equation of motion (lower path).
}
\label{fig:Schematic}
\end{figure}

The calculation of the tunneling rate from the false to true vacua was formulated 
in Refs.~\cite{Coleman:1977aa,Callan:1977aa}.
The rate is estimated by the saddle-point configuration in the Euclidean path integral called ``bounce."
Let us consider the tunneling of a scalar field $\phi$ from the false vacuum $\phi_+$ to the true vacuum $\phi_-$
(see also Fig.~\ref{fig:V}, in which $\phi_+ = 0$ and $\phi_- = 1$).
After taking the $O(d)$ symmetry of the relevant configuration into account~\cite{Coleman:1978aa,Lopes:1996aa,Byeon:2008aa,Blum:2017aa},
where $d$ denotes the number of relevant spacetime dimensions in the Euclidean space,
the equation of motion becomes
\begin{align}
\phi'' + \frac{d - 1}{r} \phi' - \frac{dV}{d\phi} 
&= 0,
\end{align}
with the boundary condition $\phi'(r = 0) = 0$ and $\phi(r = \infty) = \phi_+$.
Here the prime denotes the derivative with respect to the Euclidean radial coordinate $r$.
In order to get the bounce action, we integrate the obtained configuration 
\begin{align}
S_d
&= 
\int dr~
s_{d - 1}r^{d-1}
\left[
\frac{1}{2} \phi'^2 + V(\phi) - V(\phi_+)
\right],
\end{align}
where $s_{d - 1}$ is the surface area of $d - 1$ dimensional unit hypersphere.
The tunneling rate $\Gamma$ is dominantly determined by this bounce action
since the former is exponentially dependent on the latter: $\Gamma \propto e^{-S_d}$.
This procedure is denoted by the upper path of Fig.~\ref{fig:Schematic}.

As long as one is interested in calculating the bounce action once and for all,
iterating the above procedure changing the initial condition for $\phi$ 
(so-called overshoot/undershoot method) may be enough.
However, in many practical situations, one is often required to perform such calculations many times
and the resulting computational cost sometimes becomes huge.\footnote{
There are a number of approaches to this problem:
for example,
piecewise-linear (or polygonal) approximation~\cite{Duncan:1992aa,Dutta:2012aa,Guada:2018aa};
improved action~\cite{Kusenko:1995aa,Moreno:1998aa};
path deformation~\cite{Cline:1999aa,Wainwright:2012aa};
dampling injection~\cite{Konstandin:2006aa};
multiple shooting~\cite{Masoumi:2017ab};
auxiliary potential~\cite{Espinosa:2018aa}.
It is also possible to give lower or upper bounds on the bounce action~\cite{Dasgupta:1997aa,Sarid:1998aa,Aravind:2014aa,Masoumi:2017aa,Sato:2018aa}.
}\footnote{
One example is when we estimate the cosmological implications of thermal first-order phase transitions,
in which case we have to estimate the bounce action for various values of the cosmic temperature
in order to know the typical transition time.
}

In this letter we approach this problem with machine learning.
The main point is summarized in the lower path of Fig.~\ref{fig:Schematic}.
If one would like to know only the bounce action, and is not interested in the intermediate process,
the calculation can be regarded as a procedure to associate the potential with a number (bounce action).
For a good machine, 
this procedure may not have to be done via the bounce equation of motion,
just as a good discriminator of pictures of cats and dogs
does not (probably) recognize them in the same way as humans do.
In this sense the problem can be regarded as a usual image recognition problem.\footnote{
Such a bypassing has already been discussed in the literature.
For example, in Ref.~\cite{Mills:2017aa} the authors developed a machine 
which calculates the ground state energies of two-dimensional systems
without referring to the Schr\"odinger equation.
}

In the following sections we investigate this possibility
by training neural network with supervised learning.
We restrict ourselves to $d = 4$, but generalization to other dimensions is straightforward. 
Also, we consider only one-dimensional scalar field potential $V(\phi)$, 
and leave generalizations to multi-dimensional cases for future work.

\section{Analysis}
\label{sec:Analysis}

\subsection{Data set}
\label{subsec:Data}

We first explain the data making process.
We can safely set $V = 0$ at the false vacuum $\phi = 0$ without losing generality.
Also, since we can factor out the horizontal and vertical scaling dependence,
we impose $V = -1$ at the true vacuum $\phi = 1$.

In the analysis below we use three different classes of potentials,
each of which consists of the following polynomials:
\begin{align}
{\rm Class~1~} (C_1):~
V(\phi)
&= 
\sum_{n = 1}^7 a^{(1)}_n \phi^{n + 1},
\label{eq:C1}
\\
{\rm Class~2~} (C_2):~
V(\phi)
&= 
\sum_{n = 1}^7 a^{(2)}_n \phi^{2n},
\label{eq:C2}
\\
{\rm Class~3~} (C_3):~
V(\phi)
&= 
a^{(3)}_1 \phi^2 
+ 
\sum_{n = 2}^7 a^{(3)}_n \phi^{2n - 1}.
\label{eq:C3}
\end{align}
Here the coefficients $\left\{ a^{(1)}_n \right\}$, $\left\{ a^{(2)}_n \right\}$, and $\left\{ a^{(3)}_n \right\}$ are 
generated from random seeds $(V_{\rm max}, \phi_0, \phi_{1-}, \phi_{1+}, \phi_2)$.
Also, we impose the following for $0 \leq \phi \leq 1$:
\begin{itemize}
\item
$V$ takes a local maximum only at $\phi = \phi_0$.
\item
$V$ takes local minima only at $\phi = 0$ and $\phi = 1$.
\end{itemize}
We generate $10,000$ potentials for each class from this generating rule,
and calculate the bounce action $S_4$ with the overshoot/undershoot method.
We explain the generating process in detail in Appendix~\ref{app:Data},
but the most important point is that the potentials in these classes are made from different generating rules.
This helps us to check the performance of the machine later.

\begin{figure}
\centerline{
\includegraphics[width=0.32\columnwidth]{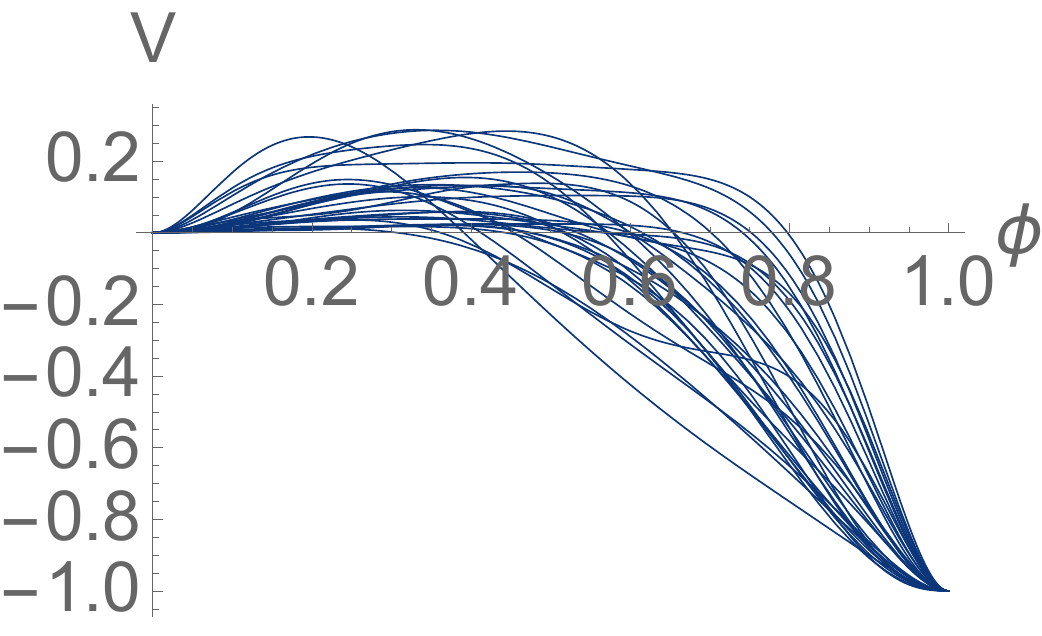}
\includegraphics[width=0.32\columnwidth]{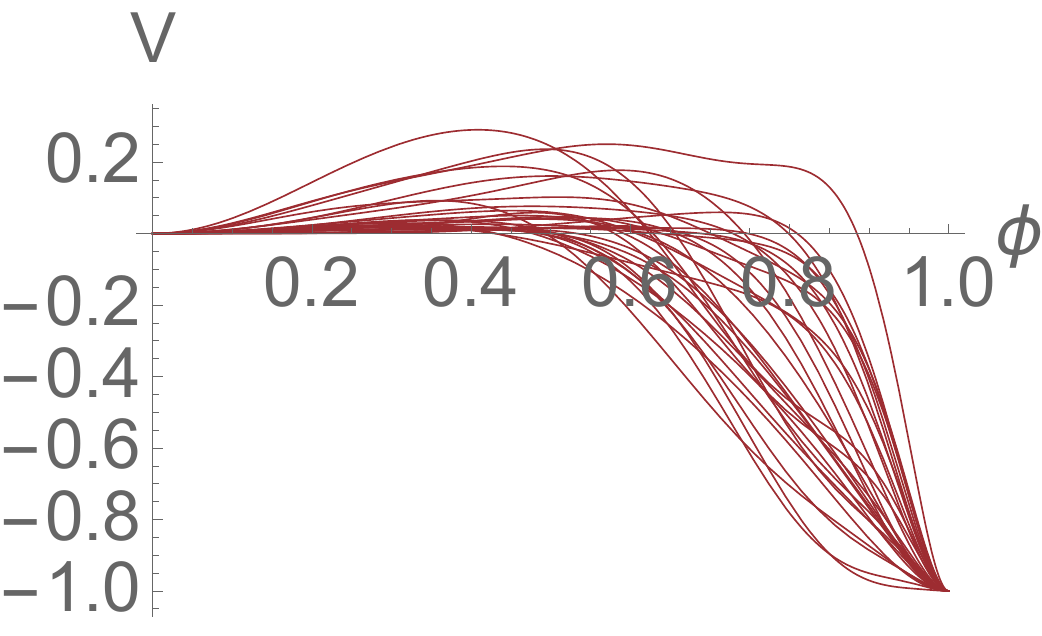}
\includegraphics[width=0.32\columnwidth]{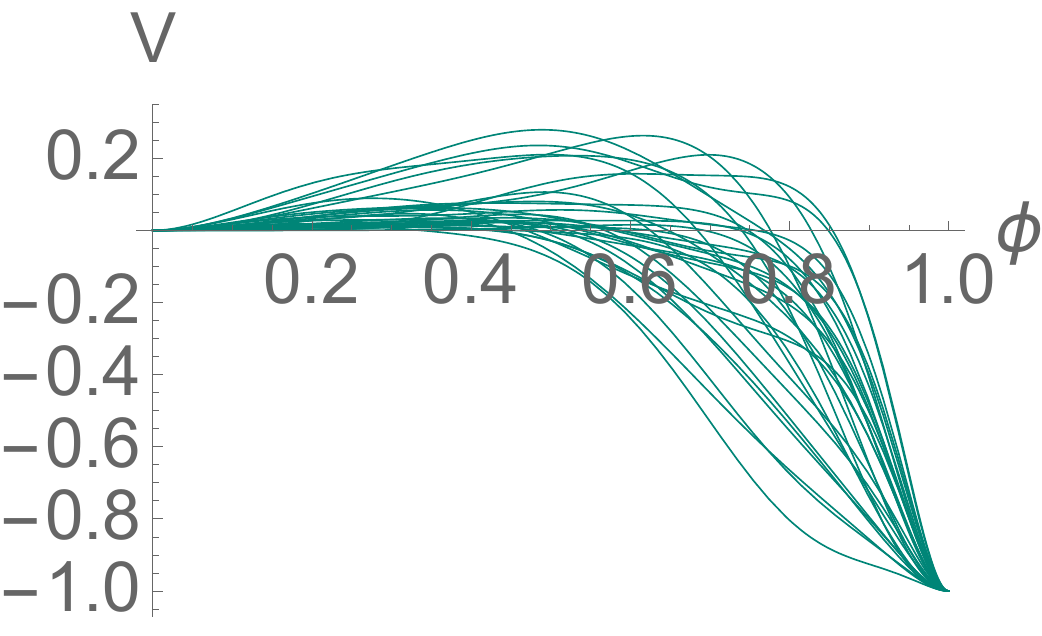}
}
\caption {\small
Randomly generated 30 potentials $V(\phi)$ in Class $C_1$--$C_3$ from left to right.
}
\label{fig:V}
\centerline{
\includegraphics[width=0.32\columnwidth]{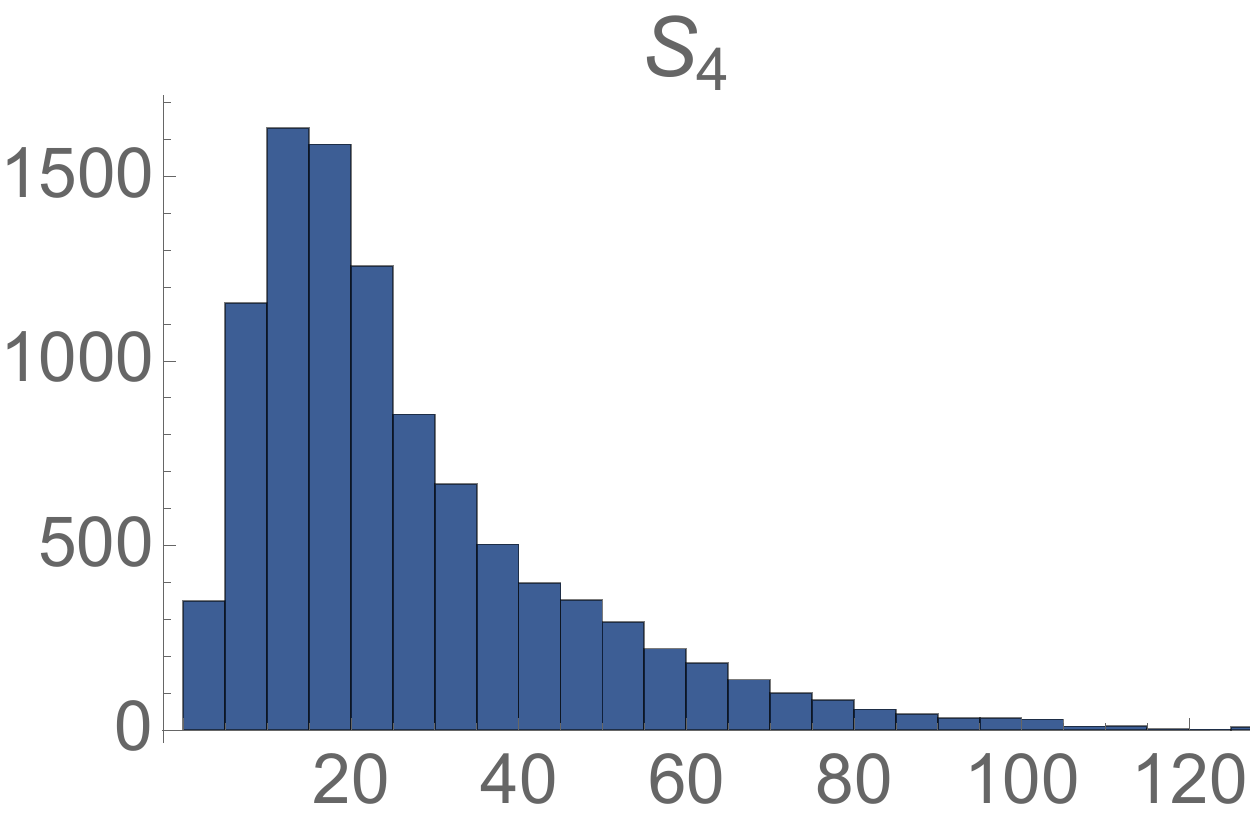}
\includegraphics[width=0.32\columnwidth]{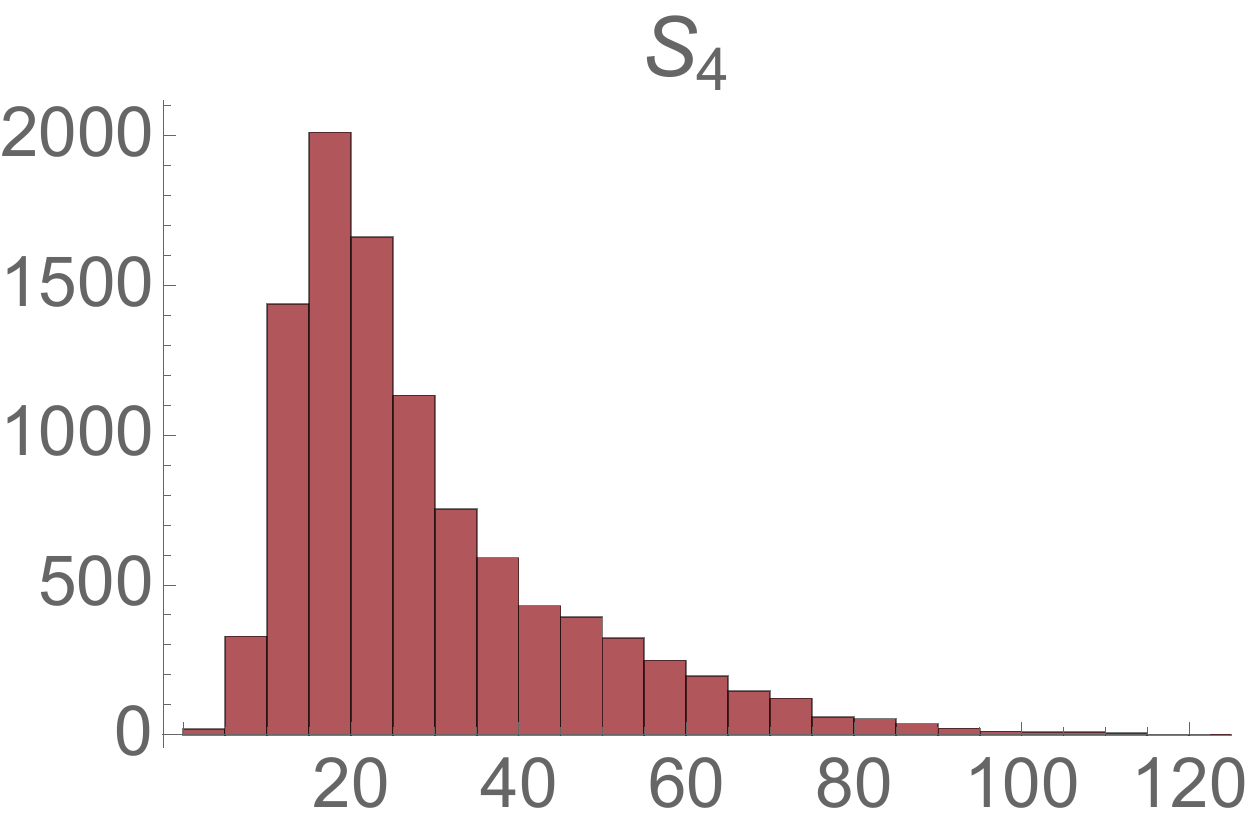}
\includegraphics[width=0.32\columnwidth]{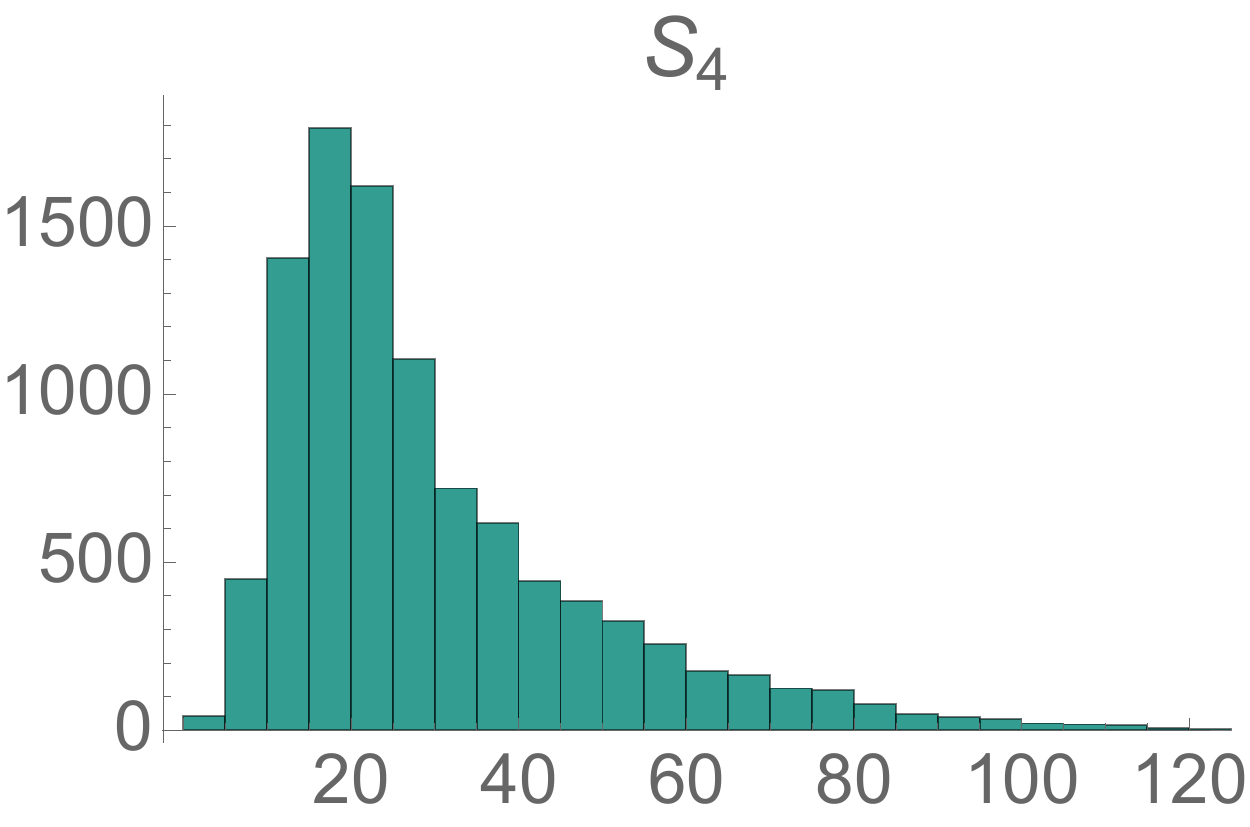}
}
\caption {\small
Distribution of the bounce action $S_4$ in Class $C_1$--$C_3$.
}
\label{fig:SHist}
\end{figure}

In Fig.~\ref{fig:V} we plot some of the potentials generated randomly for each class.
Also, in Fig.~\ref{fig:SHist} we show the distribution of the bounce action $S_4$.

In the following analysis, we use combinations of these data as ``training $\&$ test" dataset and ``application" dataset
(see Sec.~\ref{subsec:Results}).
The former is further divided into training and test datasets with a constant ratio $0.8 : 0.2$.
The neural network is trained on the training dataset, and monitored with the test dataset to avoid possible overfitting.
After training, the neural network is applied to the application dataset.

\subsection{Setup of the neural network}
\label{subsec:Setup}

\begin{figure}
\centerline{
\includegraphics[width=\columnwidth]{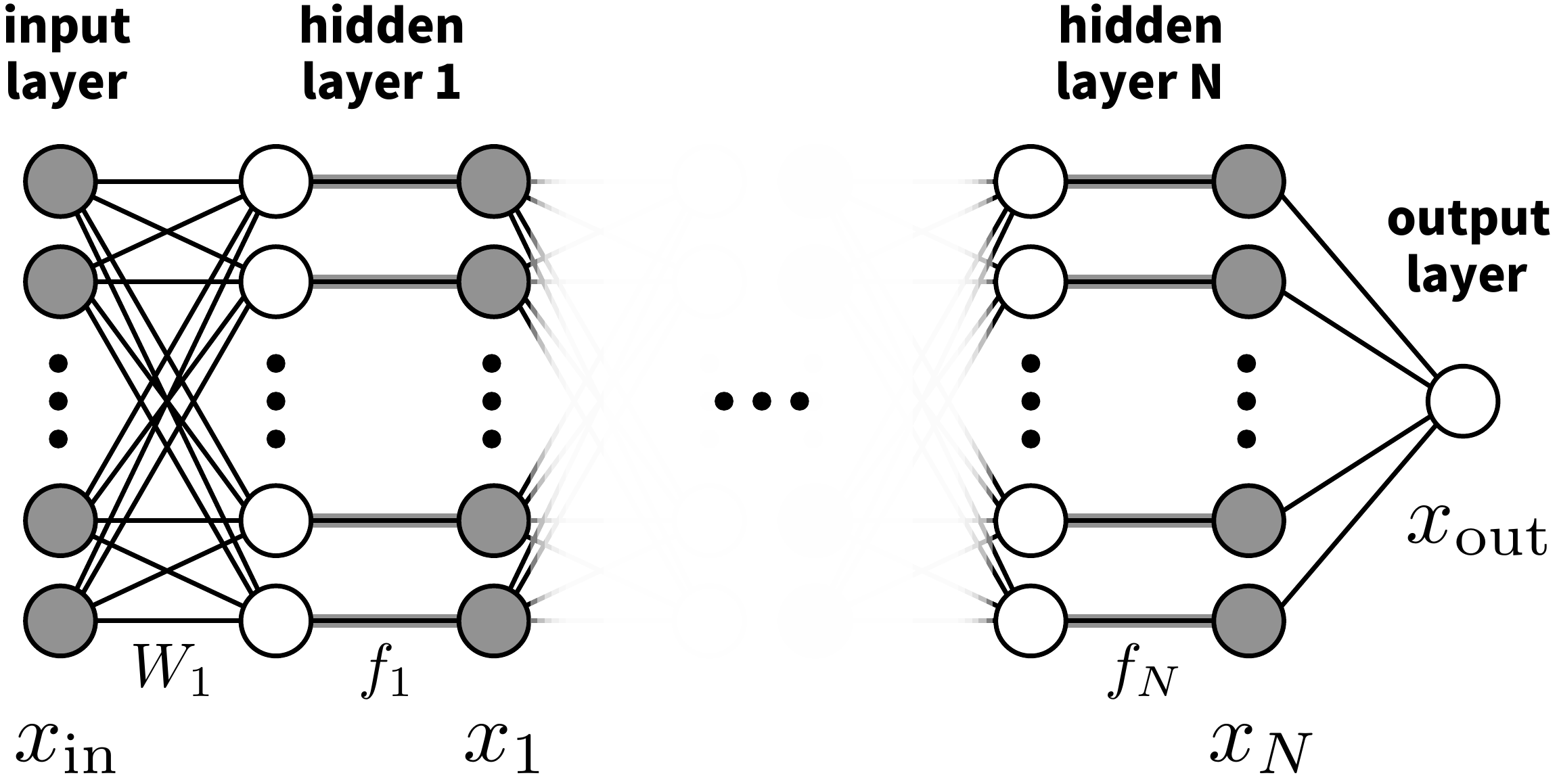}
}
\caption {\small
Schematic diagram of the neural network used in the present letter.
}
\label{fig:NN}
\end{figure}

Now we turn to the setup of the neural network.
We use a simple network schematically described in Fig.~\ref{fig:NN}.
The input data $x_{\rm in}$ consists of the values of the potential $V$ and its derivatives
\begin{align}
x_{\rm in}
= 
&\left\{
V(\phi_{\rm sample})~\left|~
\phi_{\rm sample} = \frac{1}{16},~\frac{2}{16}, \cdots,~\frac{15}{16}
\right.
\right\}
\nonumber \\
&\oplus
\left\{
\frac{dV}{d\phi}(\phi_{\rm sample})~\left|~
\phi_{\rm sample} = \frac{1}{16},~\frac{2}{16}, \cdots,~\frac{15}{16}
\right.
\right\}
\nonumber \\
&\oplus
\left\{
\frac{d^2V}{d\phi^2}(\phi_{\rm sample})~\left|~
\phi_{\rm sample} = \frac{0}{16},~\frac{1}{16}, \cdots,~\frac{16}{16}
\right.
\right\}.
\end{align}
Here we do not include $\phi_{\rm sample} = 0/16$ and $16/16$ for $V$ and $dV/d\phi$
because the values are the same among all the potentials realized in $C_1$--$C_3$.
Also, when we actually input $x_{\rm in}$ in the network,
we normalize it to make the learning more efficient
(i.e. each element in $x_{\rm in}$ is normalized as
$(x_{\rm in})_i \to ((x_{\rm in})_i - \left< (x_{\rm in})_i \right>) / \sigma_{(x_{\rm in})_i}$,
where the mean $\left< (x_{\rm in})_i \right>$ and the variance $\sigma^2_{(x_{\rm in})_i}$ 
are calculated over the training $\&$ test dataset).

For the hidden layers, we use only $N = 2$. 
The numbers of neurons are $20$ and $10$ for the first and second hidden layers, respectively.
The connection among the layers is the same as the typical neural network system:
\begin{align}
x_1
&= f_1(W_1 x_{\rm in} + b_1),
\\
x_n
&= f_n(W_n x_{n - 1} + b_n)~~(2 \leq n \leq N),
\\
x_{\rm out}
&= W_{\rm out} x_N + b_{\rm out},
\end{align}
where $W$'s and $b$'s are the weight matrices and biases, respectively.
Note that these equations should be understood in a matrix sense:
for example, $W_1$ is a $(20, 47)$ dimensional matrix while 
$x_{\rm in}$ and $b_1$ are $47$ and $20$ dimensional vectors, respectively.
Also, $f$'s are nonlinear functions known as activation functions,
which operate componentwise, e.g. $(x_m)_i = f_m ( (W_m)_{ij} (x_{m - 1})_j + (b_m)_i)$.
We adopt ReLU (rectified linear unit) for $f$'s.

In this letter we use supervised machine learning.
We train the neural network so that it makes the output value $x_{\rm out}$  
as close as possible to the true value $x_{\rm out}^{(\rm true)}$,
which is taken to be the logarithmic bounce action 
$\ln S_4$ normalized over the training $\&$ test dataset
(i.e.
$x_{\rm out}^{(\rm true)} = (\ln S_4 - \left< \ln S_4 \right>) / \sigma_{\ln S_4}$).
The weights and biases are initialized randomly at the beginning,
and the training proceeds with the update of these quantities.
We use an optimization method Adam~\cite{Kingma:2014aa} for this updating process
with the $L^1$-norm as the error function $E$:
\begin{align}
E
&=
\frac{1}{(\# {\rm~of~data})}
\sum_{\rm data} 
\left|
x_{\rm out} - x_{\rm out}^{(\rm true)}
\right|.
\end{align}
Here we do not include any regularization term in the error function
(and also we do not use any dropout~\cite{Hinton:2012aa,Srivastava:2014aa} in the training process)
because we observed no significant overfitting with the above simple setup (see also Appendix~\ref{app:Loss}).

In the training process, 
the neural network is fed with $1/10$ of the training dataset,
and one epoch consists of $10$ times of this feeding process.
The loss function is monitored with the test dataset to avoid possible overfitting.
We stop training after $10,000$ epochs.

The above setup is implemented with TensorFlow (r1.17)~\cite{Abadi:2016aa}.

\subsection{Results}
\label{subsec:Results}

\begin{figure}
\centerline{
\includegraphics[width=0.5\columnwidth]{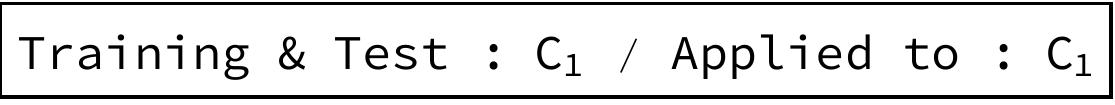}
}
\centerline{
\includegraphics[width=0.49\columnwidth]{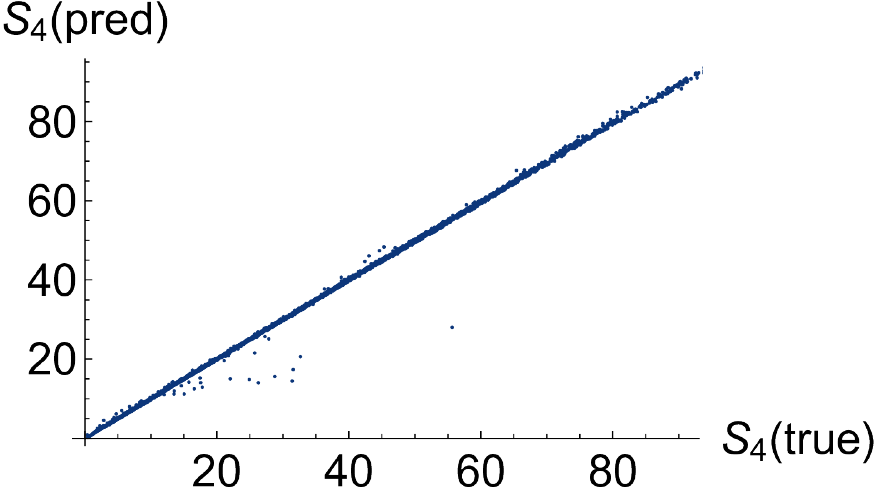}
\includegraphics[width=0.49\columnwidth]{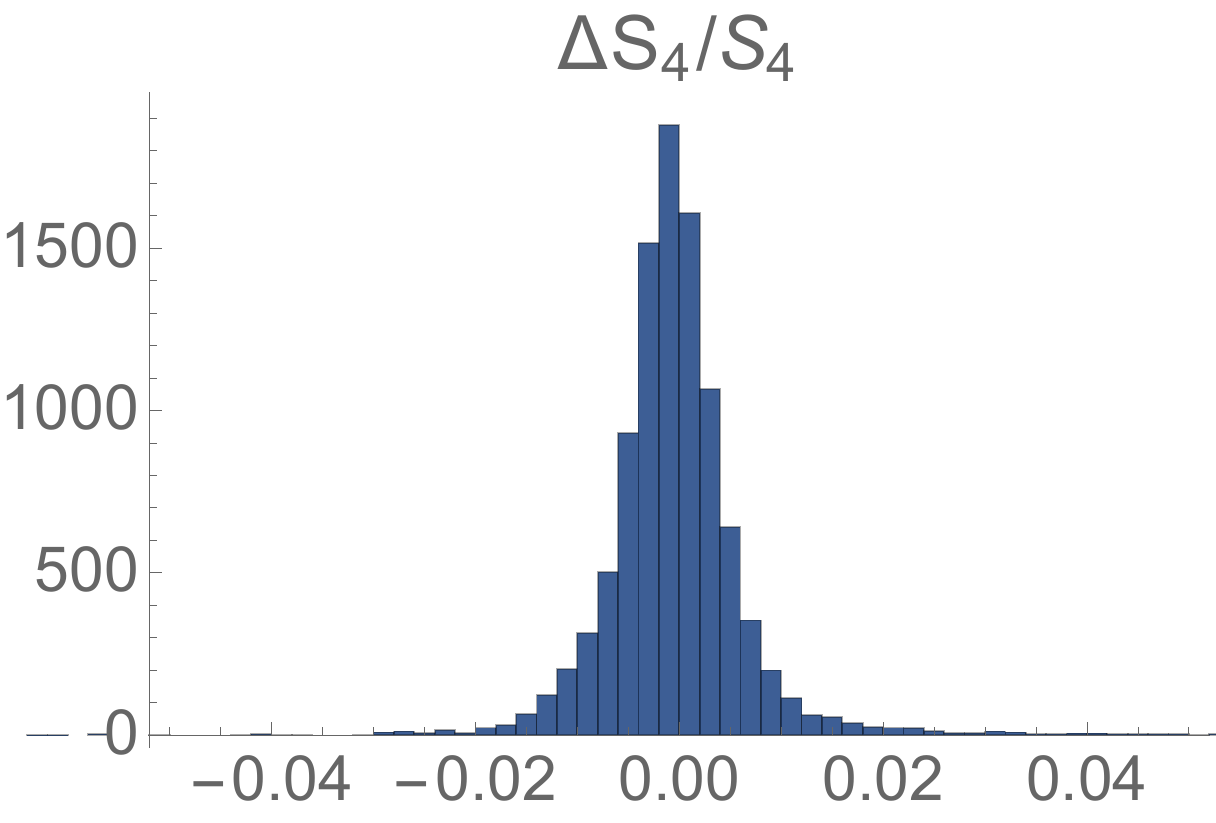}
}
\vskip 0.1in
\centerline{
\includegraphics[width=0.5\columnwidth]{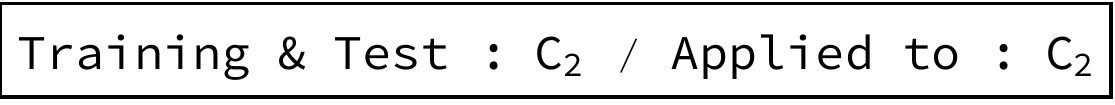}
}
\centerline{
\includegraphics[width=0.49\columnwidth]{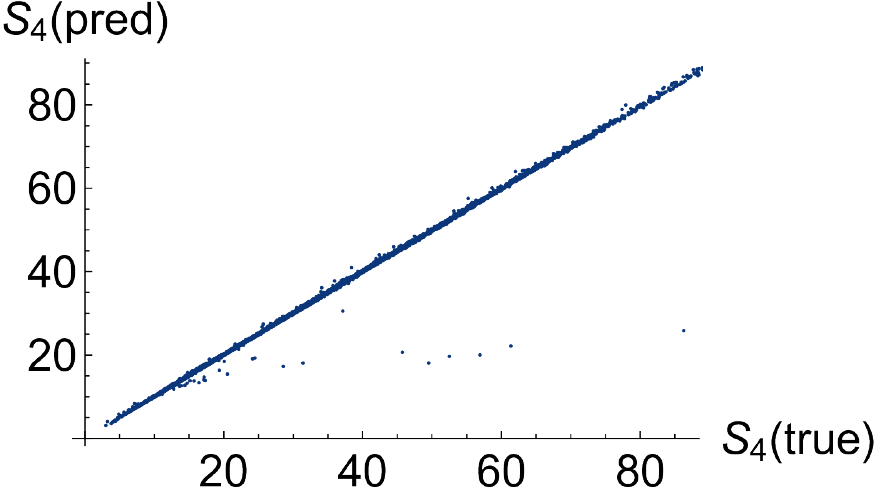}
\includegraphics[width=0.49\columnwidth]{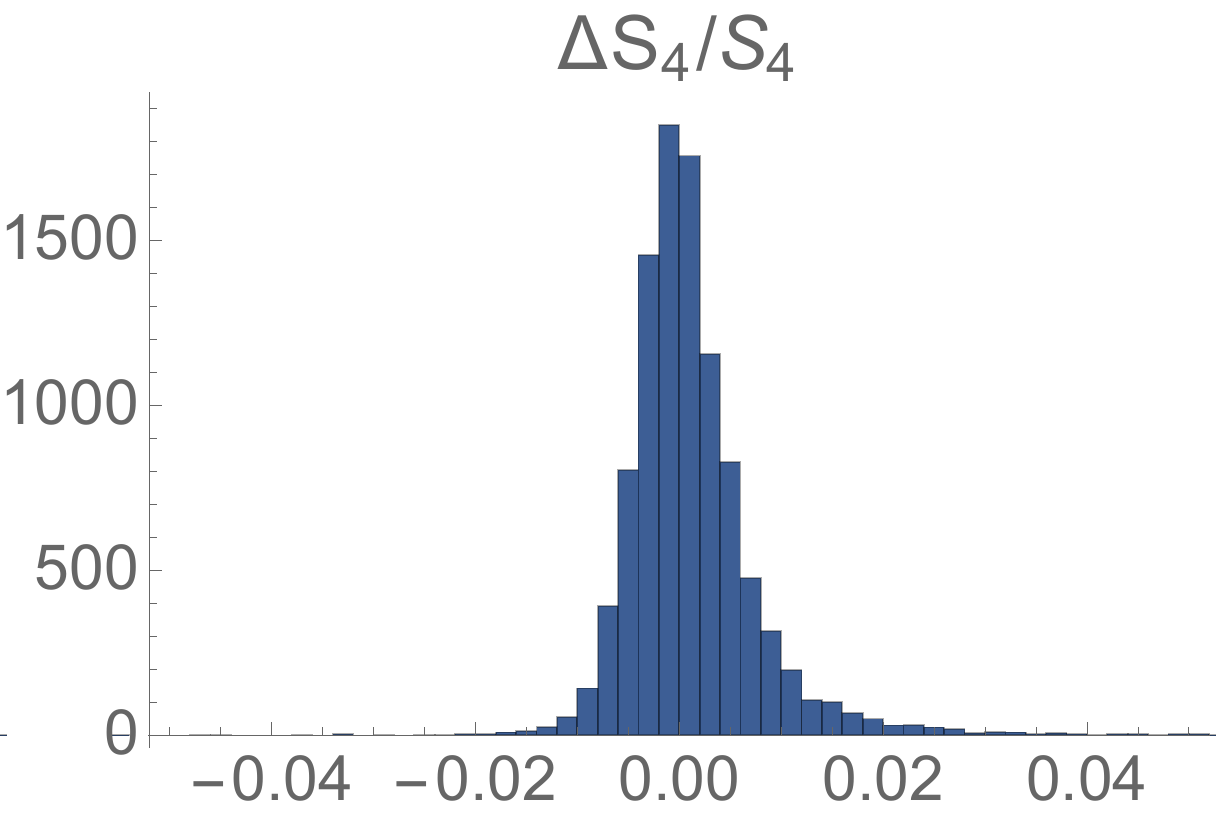}
}
\vskip 0.1in
\centerline{
\includegraphics[width=0.5\columnwidth]{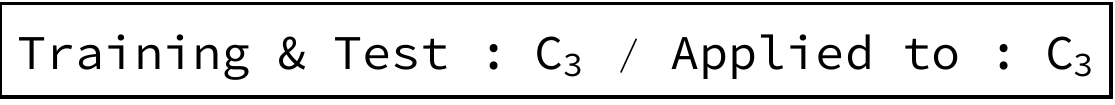}
}
\centerline{
\includegraphics[width=0.49\columnwidth]{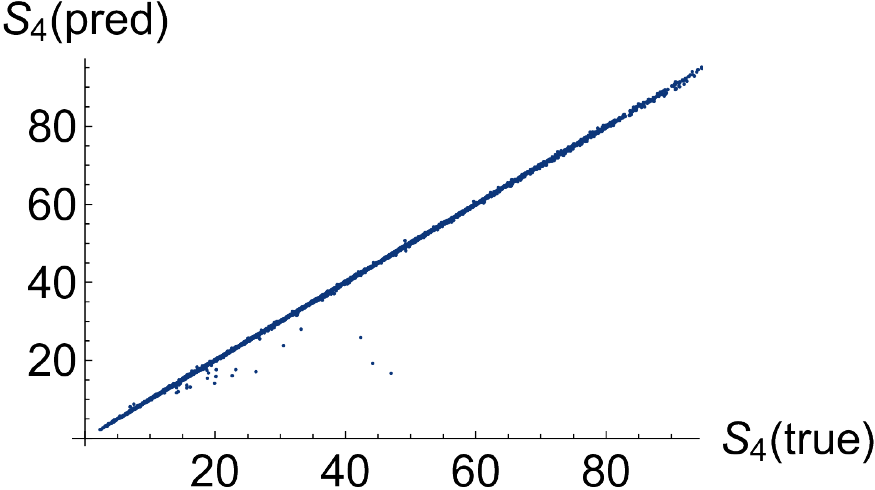}
\includegraphics[width=0.49\columnwidth]{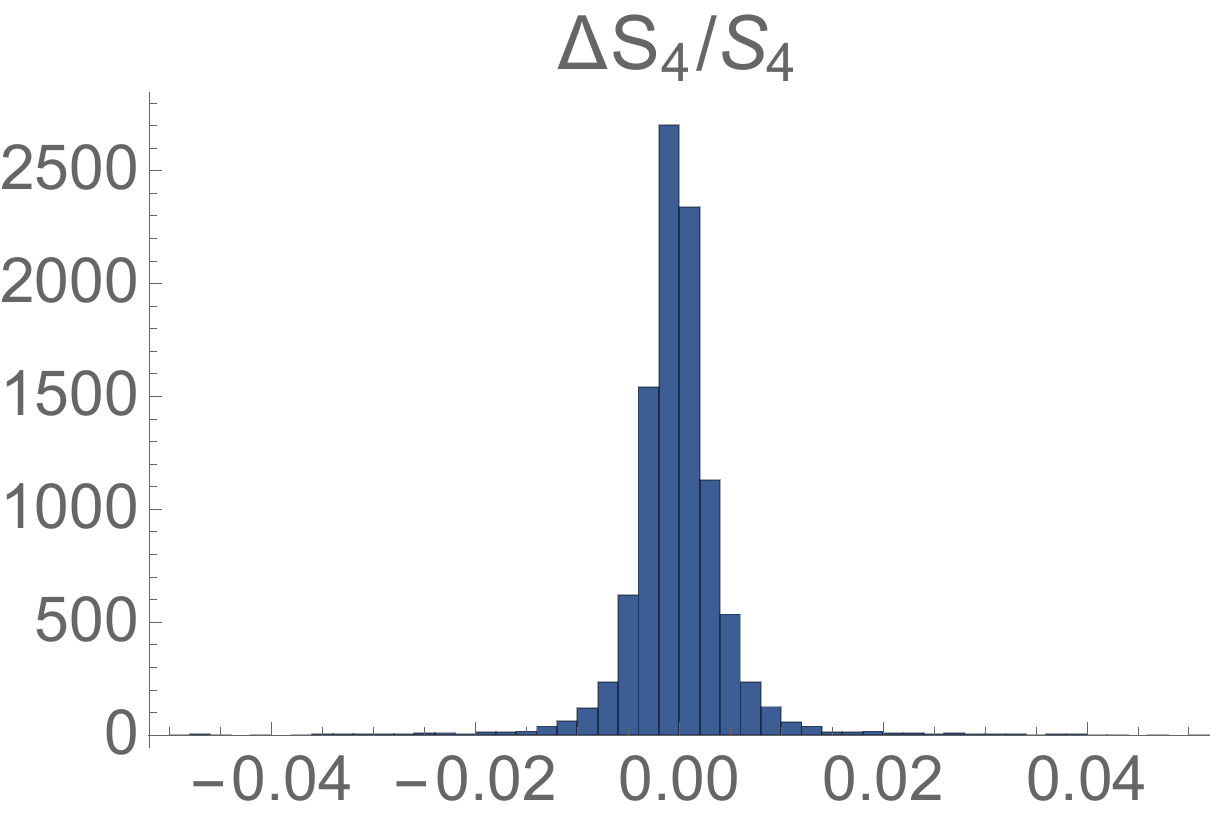}
}
\caption {\small
Scatter plot of the true value and the predicted value of $S_4$ (left),
and the distribution of the fractional error 
$\Delta S_4 / S_4 \equiv (S_4({\rm prediction}) - S_4({\rm true})) / S_4({\rm true})$ (right)
for each of $C_1$--$C_3$ from top to bottom.
}
\label{fig:S__}
\vskip 0.1in
\centerline{
\includegraphics[width=0.5\columnwidth]{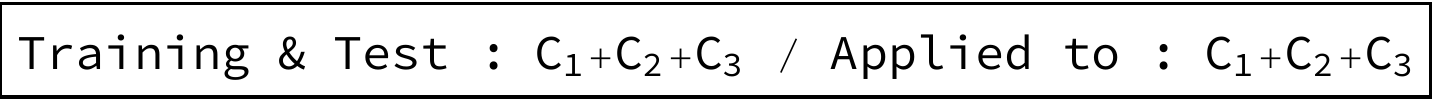}
}
\centerline{
\includegraphics[width=0.49\columnwidth]{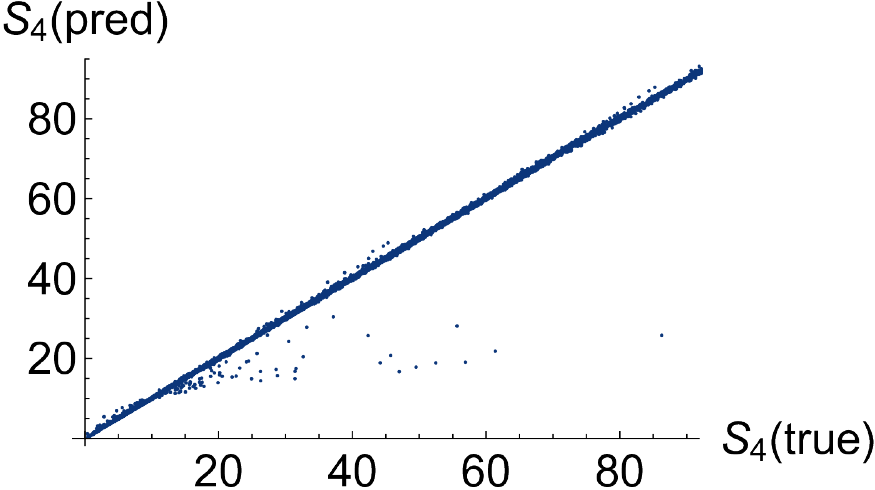}
\includegraphics[width=0.49\columnwidth]{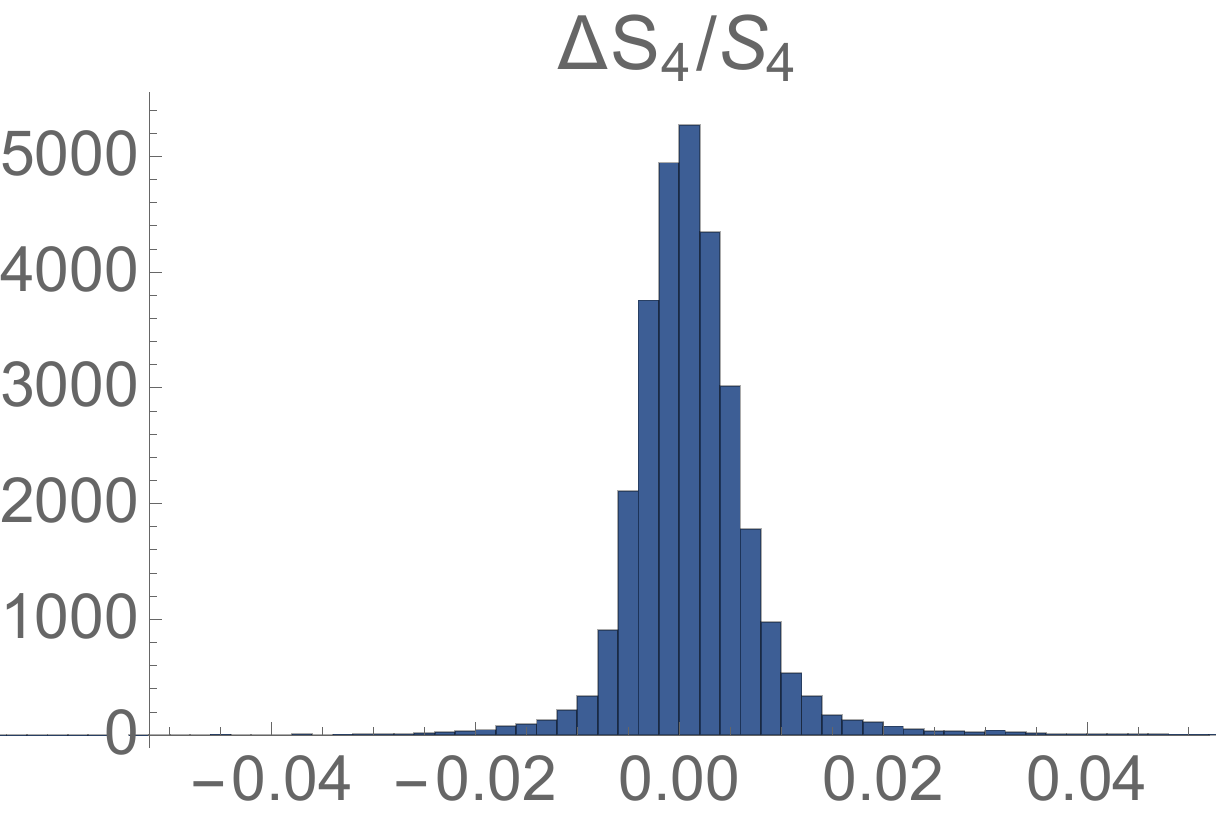}
}
\caption {\small
Same as Fig.~\ref{fig:S__}, except that $C_1 + C_2 + C_3$ is used.
}
\label{fig:S______}
\end{figure}

We first train the neural network taking each class $C_1$--$C_3$ as 
the training $\&$ test dataset {\it and}~$\;$the application dataset.
For each case, we make $10$ neural networks following the procedure described in Sec.~\ref{subsec:Setup},
and compute the resulting distribution of the predictions for $S_4$ 
and the error $\Delta S_4 / S_4 \equiv (S_4({\rm prediction}) - S_4({\rm true})) / S_4({\rm true})$ 
for the application dataset.
Fig.~\ref{fig:S__} shows the predicted values of $S_4$ (left panels) 
and the fractional error $\Delta S_4 / S_4$ (right panels) calculated by one of the $10$ neural networks selected randomly.
Also, we calculate the average $\left< |\Delta S_4 / S_4| \right>$ over the application dataset for each network
and then compute its mean 
$\left<\left< |\Delta S_4 / S_4| \right>\right>$ 
and variance
$\sigma_{\left< |\Delta S_4 / S_4| \right>}$ over the 10 networks.
The result is summarized in Table~\ref{tbl:Prec}.
It is seen that the neural networks consistently perform at a sub-percent level for each of $C_1$--$C_3$.

Next we train the network with the dataset consisting of 
$C_1$ {\it and} $C_2$ {\it and} $C_3$ 
(30,000 data in total, of which $24,000$ are used for training and $6,000$ are used to monitor possible overfitting),
in order to see how the performance becomes 
when the network is trained with a mixture of potentials made from different generating rules
(see also Appendix~\ref{app:Log}).
The result is shown in Fig.~\ref{fig:S______} and Table~\ref{tbl:Prec}.
It is seen that the neural networks still perform at a sub-percent level.

\begin{figure}
\centerline{
\includegraphics[width=0.5\columnwidth]{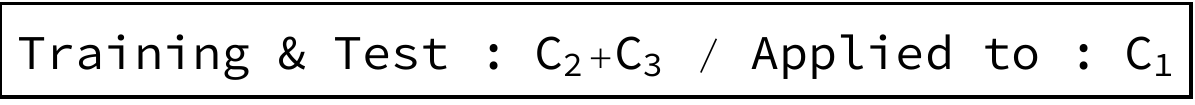}
}
\centerline{
\includegraphics[width=0.49\columnwidth]{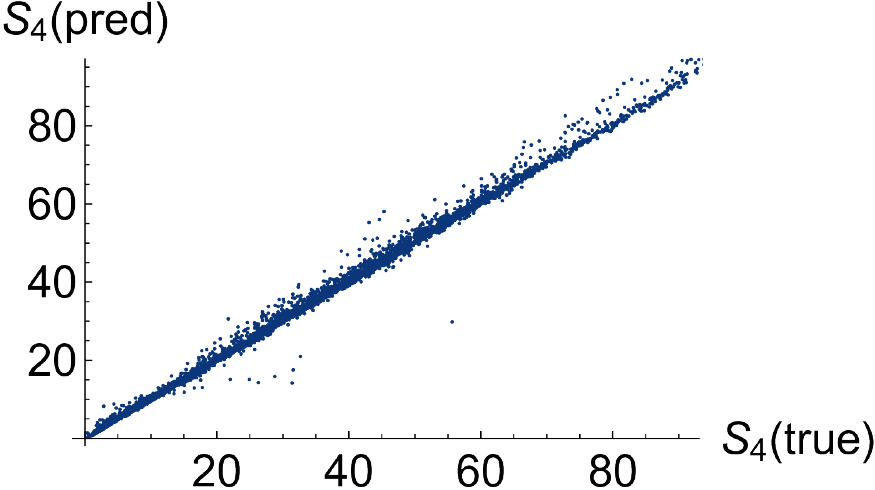}
\includegraphics[width=0.49\columnwidth]{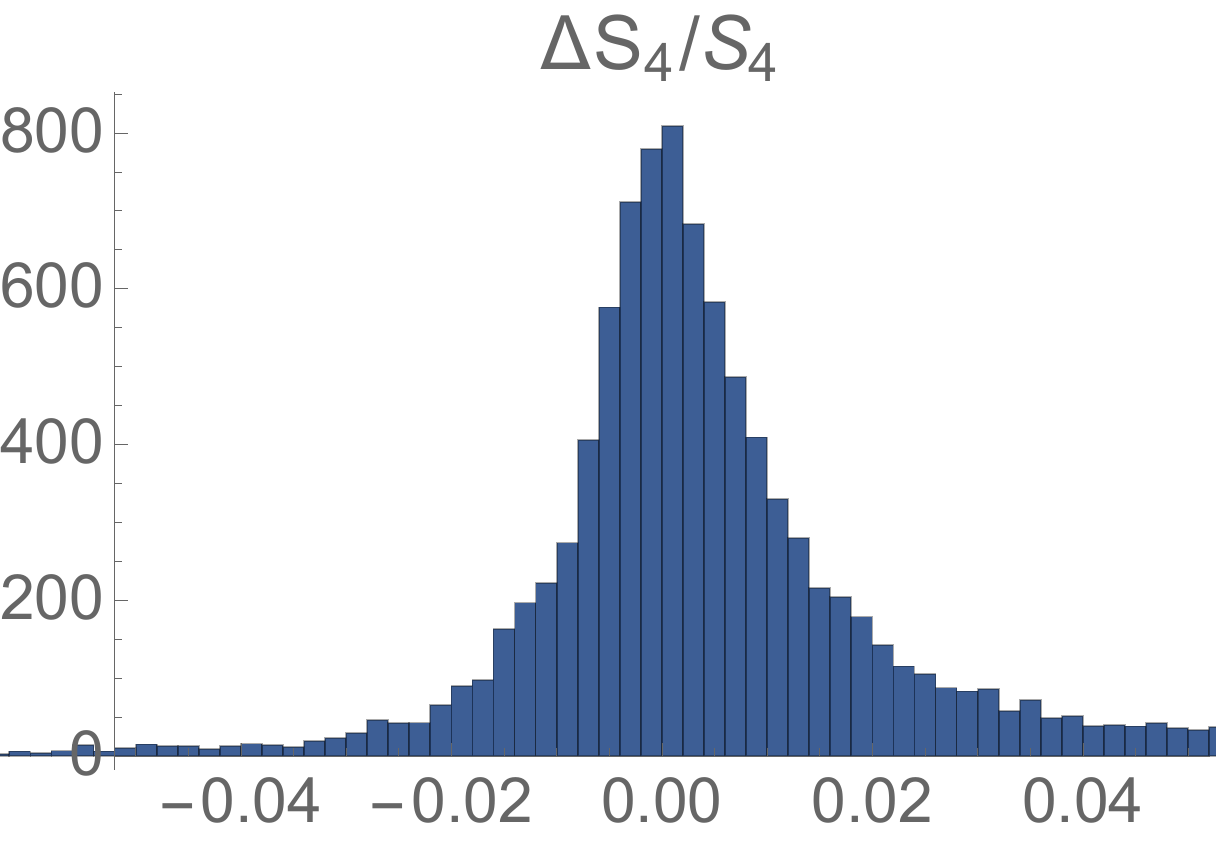}
}
\vskip 0.1in
\centerline{
\includegraphics[width=0.5\columnwidth]{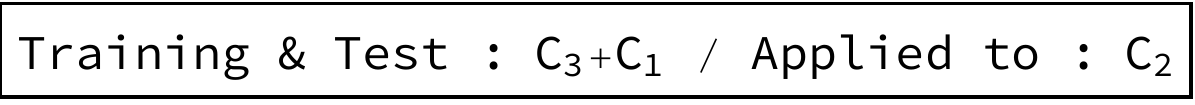}
}
\centerline{
\includegraphics[width=0.49\columnwidth]{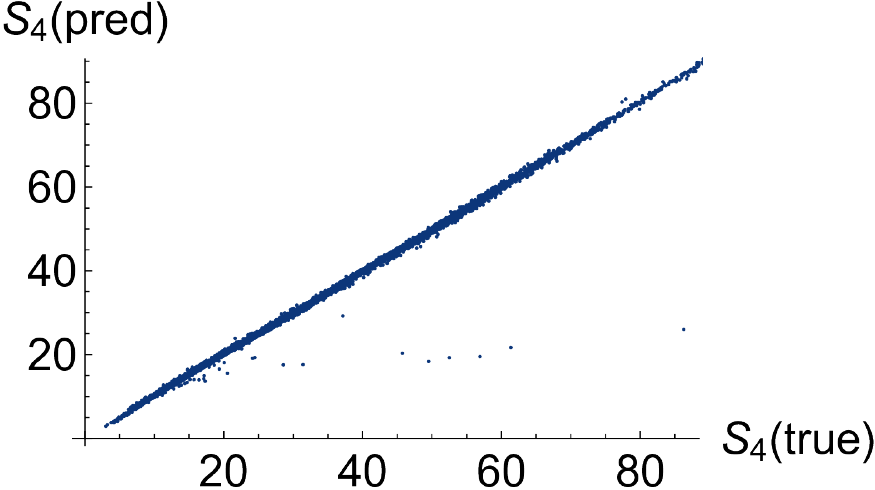}
\includegraphics[width=0.49\columnwidth]{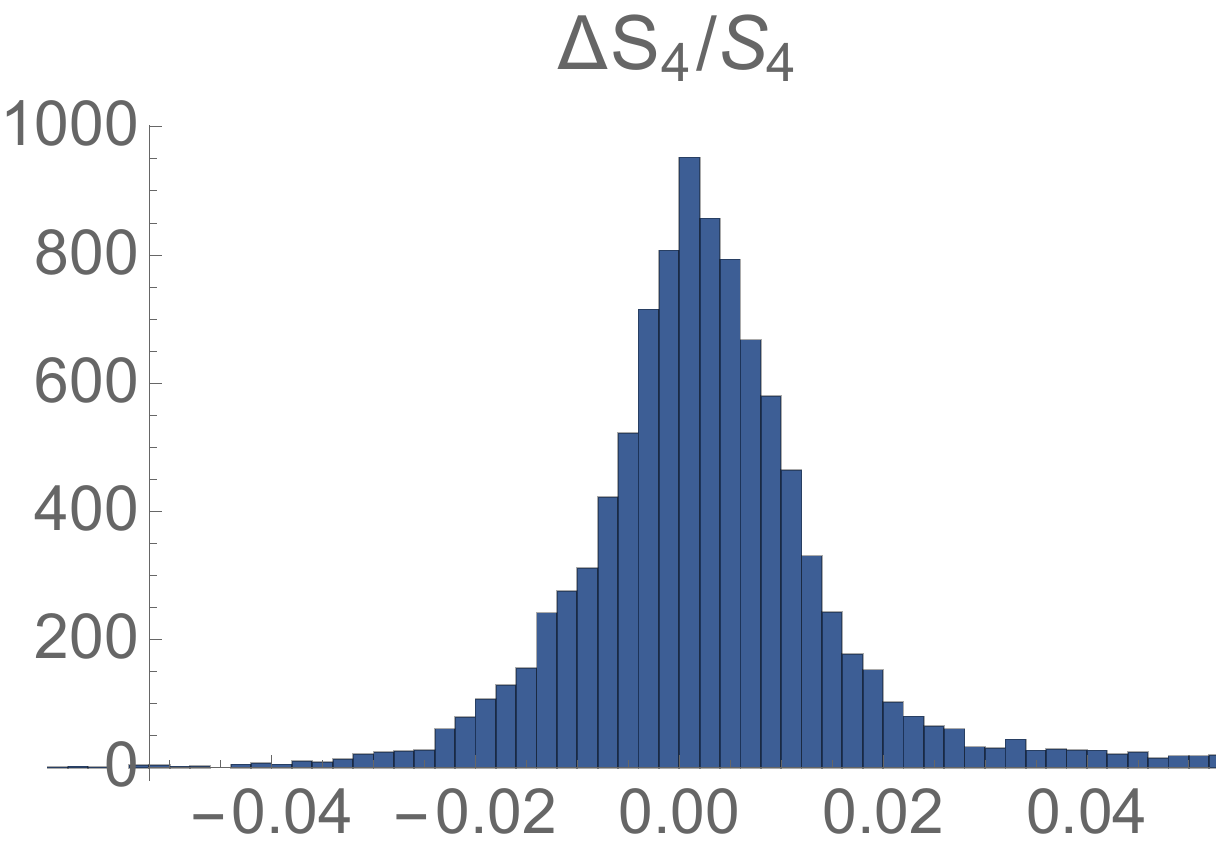}
}
\vskip 0.1in
\centerline{
\includegraphics[width=0.5\columnwidth]{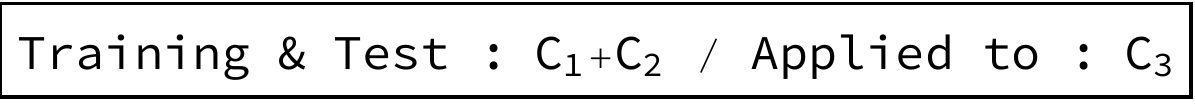}
}
\centerline{
\includegraphics[width=0.49\columnwidth]{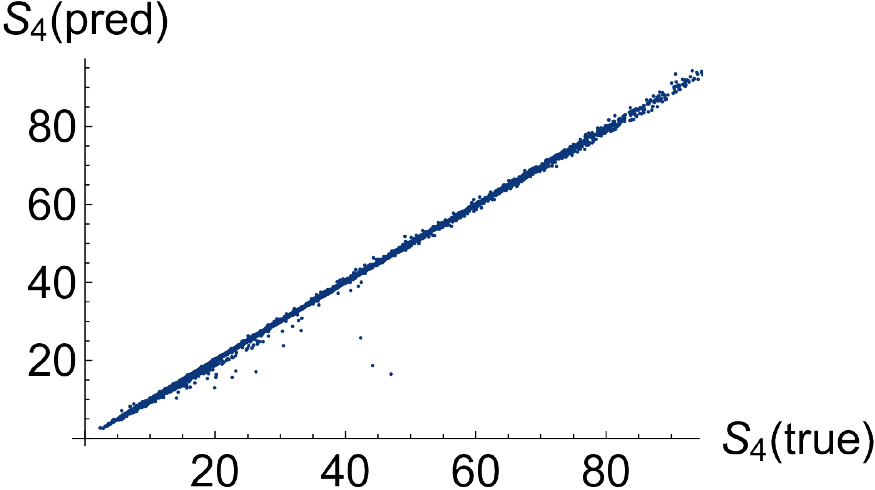}
\includegraphics[width=0.49\columnwidth]{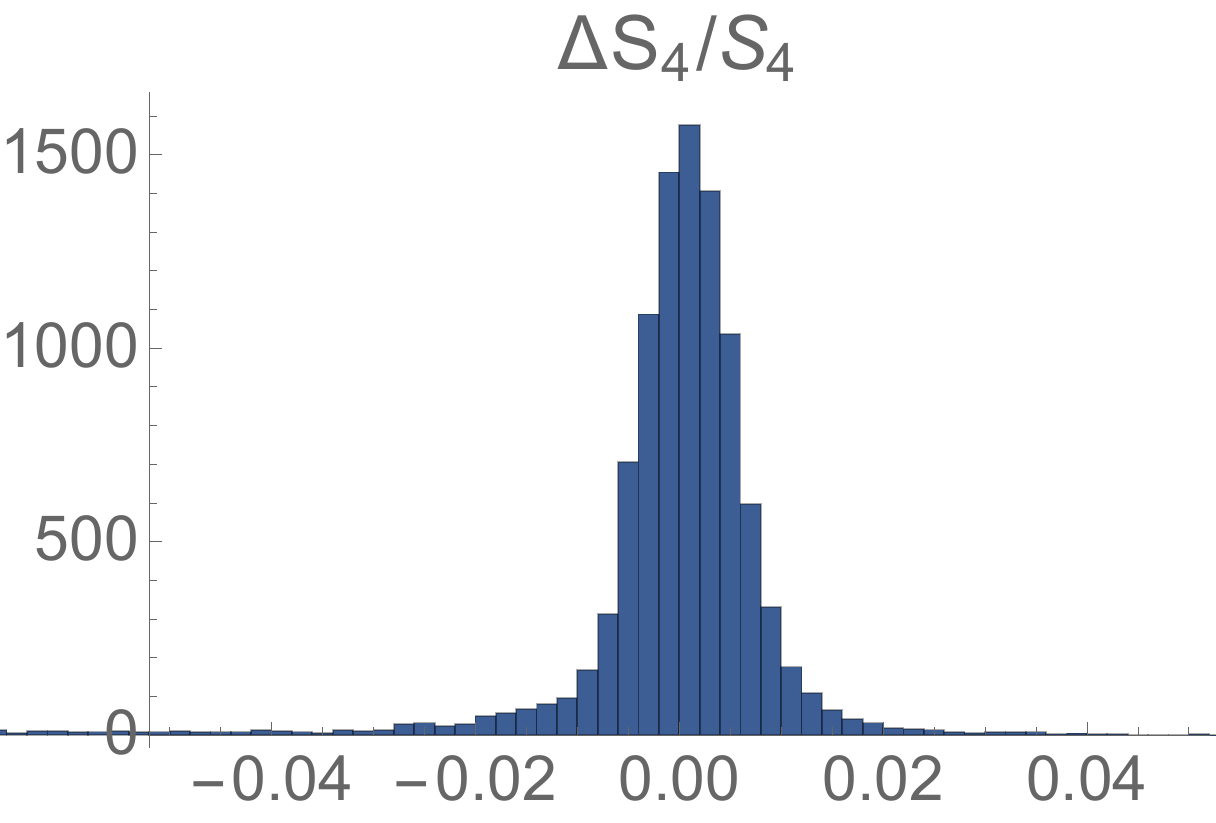}
}
\caption {\small
Same as Fig.~\ref{fig:S__}, except that $C_2 + C_3$, $C_3 + C_1$, and $C_1 + C_2$ are used.
}
\label{fig:S___}
\end{figure}

Finally we train the network with the dataset 
$C_2 + C_3$, $C_3 + C_1$, and $C_1 + C_2$,
and apply it to the dataset $C_1$, $C_2$, and $C_3$, respectively.
This is in order to check how well the neural network performs 
when they are given potentials made from a generating rule they have never seen before.
The result is summarized in Fig.~\ref{fig:S___} and Table~\ref{tbl:Prec}.
It is seen that, when the network is trained with $C_3 + C_1$ or $C_1 + C_2$, it still performs at a $1\%$ level,
while for $C_2 + C_3$ case the performance drops to a $2\%$ level.
One possible reason is as follows:
As seen from Eqs.~(\ref{eq:C1})--(\ref{eq:C3}), the classes $C_2$ and $C_3$ are ``close"
in that they use much higher polynomials than $C_1$.
This tendency can also be observed in Fig.~\ref{fig:V},
where the $\phi$ values at the zero-crossing point of $V$ tend to be smaller (larger) for $C_1$ ($C_2$ and $C_3$).
Therefore, for the neural networks trained with $C_2$ and $C_3$, 
the potentials in $C_1$ are too ``far" from what they have learned.
However, it should be noted that these networks still perform at a percent level.

In practical situations, the generating rules of the potentials in typical particle physics setups may not be that many.
Therefore, if we train the neural network with the potentials generated from such rules,
the situation will correspond to $C_1 + C_2 + C_3$ case in Table~\ref{tbl:Prec}.

\begin{table}
\caption {\small
Summary of the obtained precision of the neural network.
The quantities $\left<\left< |\Delta S_4 / S_4| \right>\right>$ 
and $\sigma_{\left< |\Delta S_4 / S_4| \right>}$ 
are the mean and variance of $\left< |\Delta S_4 / S_4| \right>$ 
calculated from the 10 networks, as explained in Sec.~\ref{subsec:Results}.
}
\begin{tabular}{|c|c|c|c|} \hline
Training $\&$ Test & Applied to & 
$\left<\left< |\Delta S_4 / S_4| \right>\right>$ & 
$\sigma_{\left< |\Delta S_4 / S_4| \right>}$ \\ \hline\hline
$C_1$ & $C_1$ & $0.00607$ & $0.00075$ \\ \hline 
$C_2$ & $C_2$ & $0.00423$ & $0.00055$ \\ \hline
$C_3$ & $C_3$ & $0.00418$ & $0.00033$ \\ \hline\hline
$C_1 + C_2 + C_3$ & $C_1 + C_2 + C_3$ & $0.00503$ & $0.00035$ \\ \hline\hline
$C_2 + C_3$ & $C_1$ & $0.0248$ & $0.0064$ \\ \hline
$C_3 + C_1$ & $C_2$ & $0.0128$ & $0.0029$ \\ \hline
$C_1 + C_2$ & $C_3$ & $0.00903$ & $0.0018$ \\ \hline
\end{tabular}
\label{tbl:Prec}
\end{table}

\section{Discussion and conclusions}
\label{sec:DC}

In this letter, we studied the calculation of the bounce action with neural network.
Our main point is summarized in Fig.~\ref{fig:Schematic}:
the standard procedure to calculate the bounce action is 
first to solve the bounce equation of motion and then integrate the solution.
However, by regarding the whole process as a number (i.e. the bounce action) 
generating process associated with the potential,
we may be able to train the machine to learn the relation
without telling the bounce equation of motion.

We illustrated this point by using a simple neural network 
trained with one-dimensional potentials generated randomly.
It was found that the neural network actually performs at a percent level (or better) 
in calculating the bounce action for one-dimensional tunneling problem (see Table~\ref{tbl:Prec}).

In this letter we considered only one-dimensional potentials.
In generalizing to multi-dimensional potentials,
state-of-the-art techniques such as convolutional neural network (CNN)~\cite{Krizhevsky:2012aa} might be helpful.
Ultimately, it is an interesting possibility to share a well-trained machine among the community,
which makes it much easier for many researchers to study the implications of first-order phase transitions.

\section*{Acknowledgment}

The author is thankful to A.~Tomiya for having nice lectures on machine learning in KIAS, Seoul.
The author is grateful to J.~Espinosa, J.~Halverson and M.~Nemev\v{s}ek for helpful comments.
This work is supported by IBS under the project code, IBS-R018-D1.

\appendix

\section{Data generating process}
\label{app:Data}

In this appendix we explain the generating process of the data used in the analysis.

As mentioned in Sec.~\ref{subsec:Data}, 
we use different types of potentials (\ref{eq:C1})--(\ref{eq:C3}) in the analysis.
The coefficients $\left\{ a^{(1)}_n \right\}$, $\left\{ a^{(2)}_n \right\}$, and $\left\{ a^{(3)}_n \right\}$  
are calculated from randomly generated seeds $(V_{\rm max}, \phi_0, \phi_{1-}, \phi_{1+}, \phi_2)$.
Here the latter quantities mean
\begin{itemize}
\item
$V$ takes a local maximum $V_{\rm max}$ at $\phi = \phi_0$.
\item
$V$ takes local minima $0$ and $-1$ at $\phi = 0$ and $\phi = 1$, respectively.
\item
$dV/d\phi$ takes a local maximum at $\phi = \phi_{1+}$.
\item
$dV/d\phi$ takes a local minimum at $\phi = \phi_{1-}$.
\item
$d^2V/d\phi^2$ takes a local minimum at $\phi = \phi_2$.
\end{itemize}
These requirements uniquely determine the coefficients $\left\{ a_n \right\}$ for each of $C_1$--$C_3$.
Here note that $V$ automatically takes a local minimum $0$ at $\phi = 0$ with the parametrization 
(\ref{eq:C1})--(\ref{eq:C3}).
We randomly generate $(V_{\rm max}, \phi_0, \phi_{1-}, \phi_{1+}, \phi_2)$ under the following conditions:
\begin{itemize}
\item
$\log_{10} V_{\rm max}$ is sampled uniformly from $[-2, -0.5]$.
\item
Four numbers are generated randomly in $[0,1]$, and 
they are identified with $\phi_{1+} < \phi_0 < \phi_2 < \phi_{1-}$ with probability $0.5$, 
while 
they are identified with $\phi_{1+} < \phi_2 < \phi_0 < \phi_{1-}$ with probability $0.5$.
\end{itemize}
From the potentials generated from these rules, we keep those which satisfy the following conditions:
\begin{itemize}
\item
$V$ takes a local maximum {\it only} at $\phi = \phi_0$ for $0 \leq \phi \leq 1$.
\item
$V$ takes local minima {\it only} at $\phi = 0$ and $\phi = 1$ for $0 \leq \phi \leq 1$.
\end{itemize}
The bounce action $S_4$ is calculated from these potentials by using the overshoot/undershoot method.

\section{Machine precision on logarithmic potential}
\label{app:Log}

\begin{figure}
\centerline{
\includegraphics[width=0.49\columnwidth]{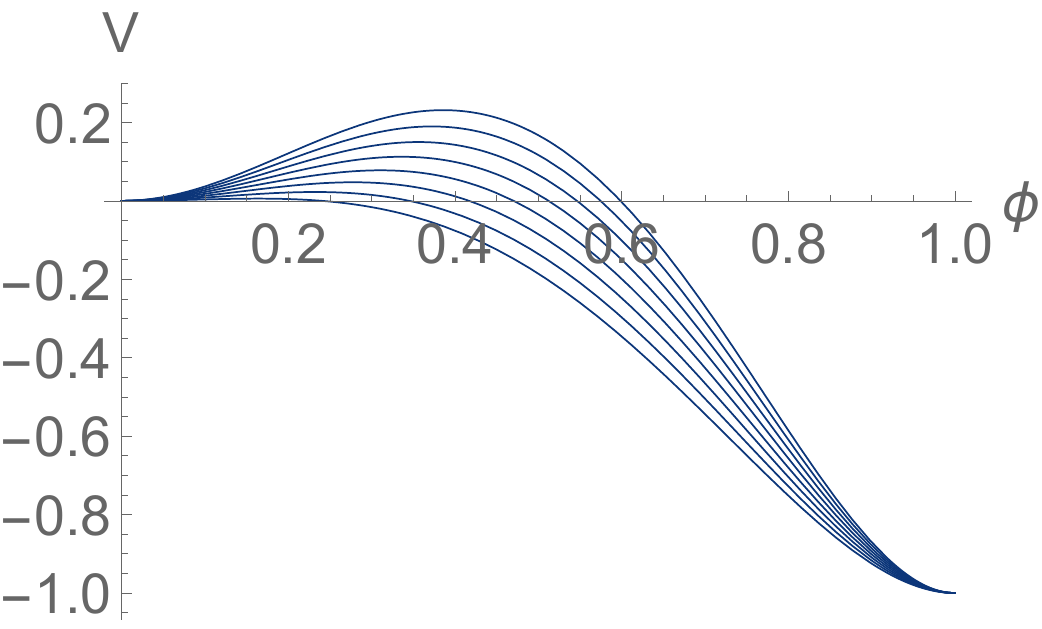}
\includegraphics[width=0.49\columnwidth]{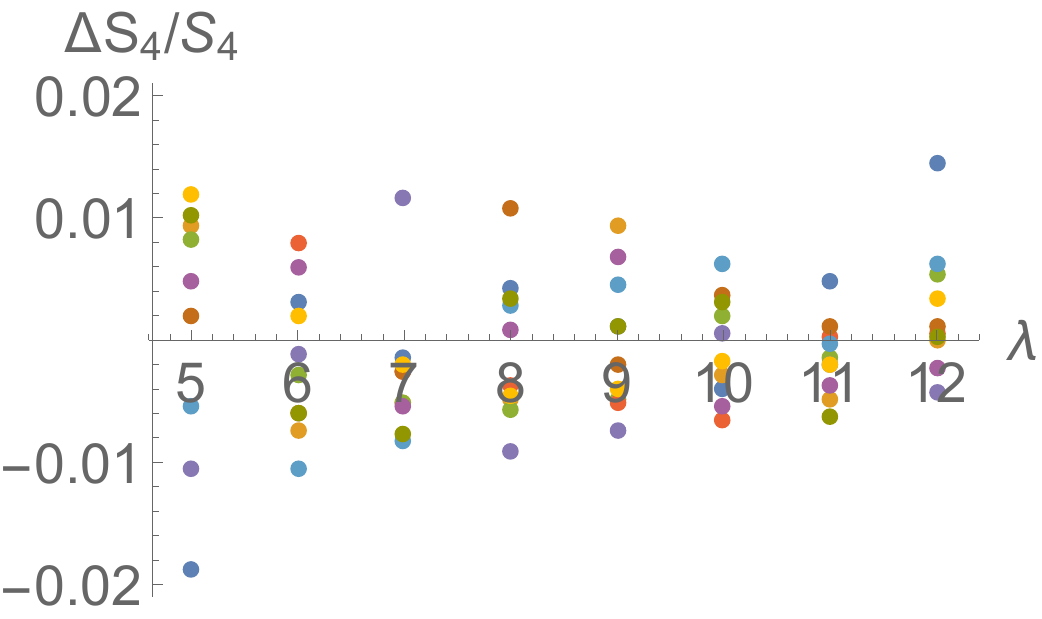}
}
\caption {\small
Potential (\ref{eq:VCW}) for $\lambda = 5, 6, \dots, 12$ from bottom to top (left).
The fractional error $\Delta S_4 / S_4$ is calculated over 
the 10 neural networks trained with $C_1 + C_2 + C_3$ dataset (right).
Different colors correspond to different neural networks.
}
\label{fig:CW}
\end{figure}

In this appendix we check the precision of our neural network with the following potential:
\begin{align}
V(\phi)
&= 
\frac{m^2}{2}\phi^2 + \frac{\lambda}{4}\log(\phi/\mu)\phi^4.
\label{eq:VCW}
\end{align}
In order for this potential to have a local minimum $V = -1$ at $\phi = 1$,
$m$ and $\mu$ are related to $\lambda$ through
\begin{align}
m^2
&= 
\frac{\lambda}{4} - 4,
~~~~
\ln \mu
= 
\frac{1}{2} - \frac{4}{\lambda}.
\end{align}
The left panel of Fig.~\ref{fig:CW} shows the potential for various values of $\lambda$.
We calculate the bounce action $S_4$ for this potential 
with the 10 neural networks trained with $C_1 + C_2 + C_3$ dataset,
and compare it with the true value calculated from the overshoot/undershoot method.
The resulting values of $\Delta S_4 / S_4$ are shown in the right panel of Fig.~\ref{fig:CW}.
It is seen that the neural networks also perform at a (sub-)percent level for this potential.

\section{Evolution of the loss function}
\label{app:Loss}

\begin{figure}
\centerline{
\includegraphics[width=\columnwidth]{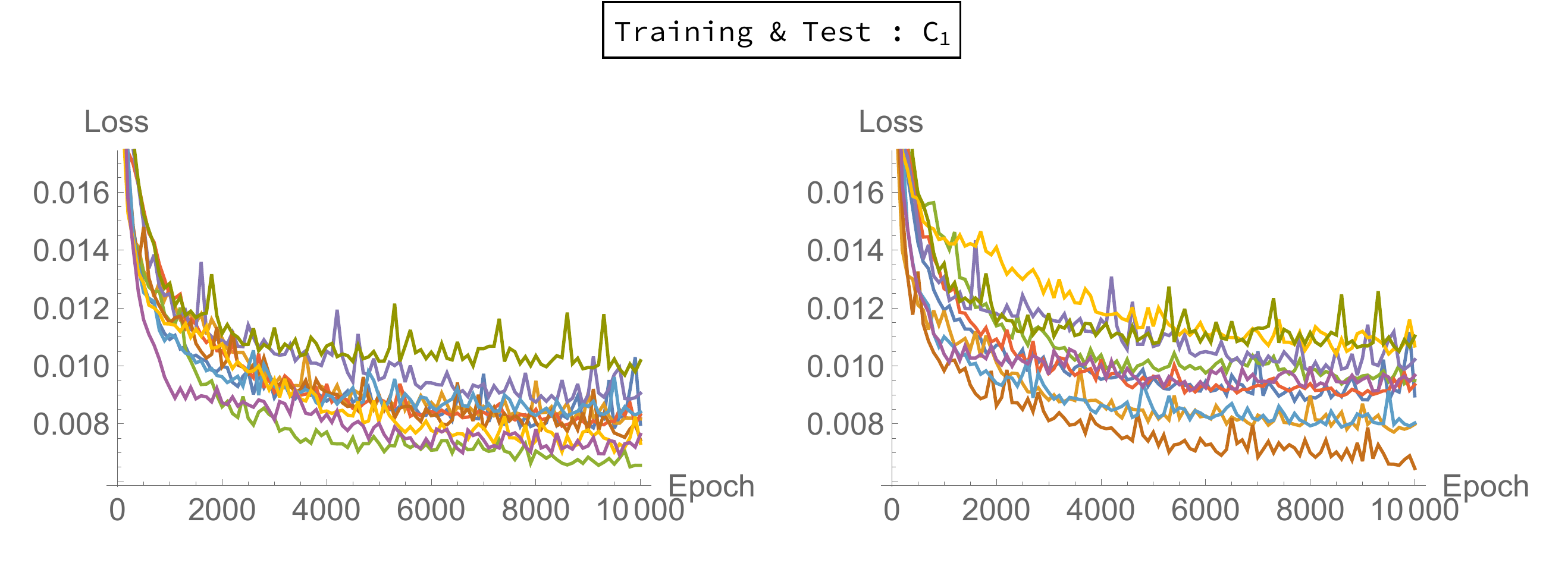}
}
\centerline{
\includegraphics[width=\columnwidth]{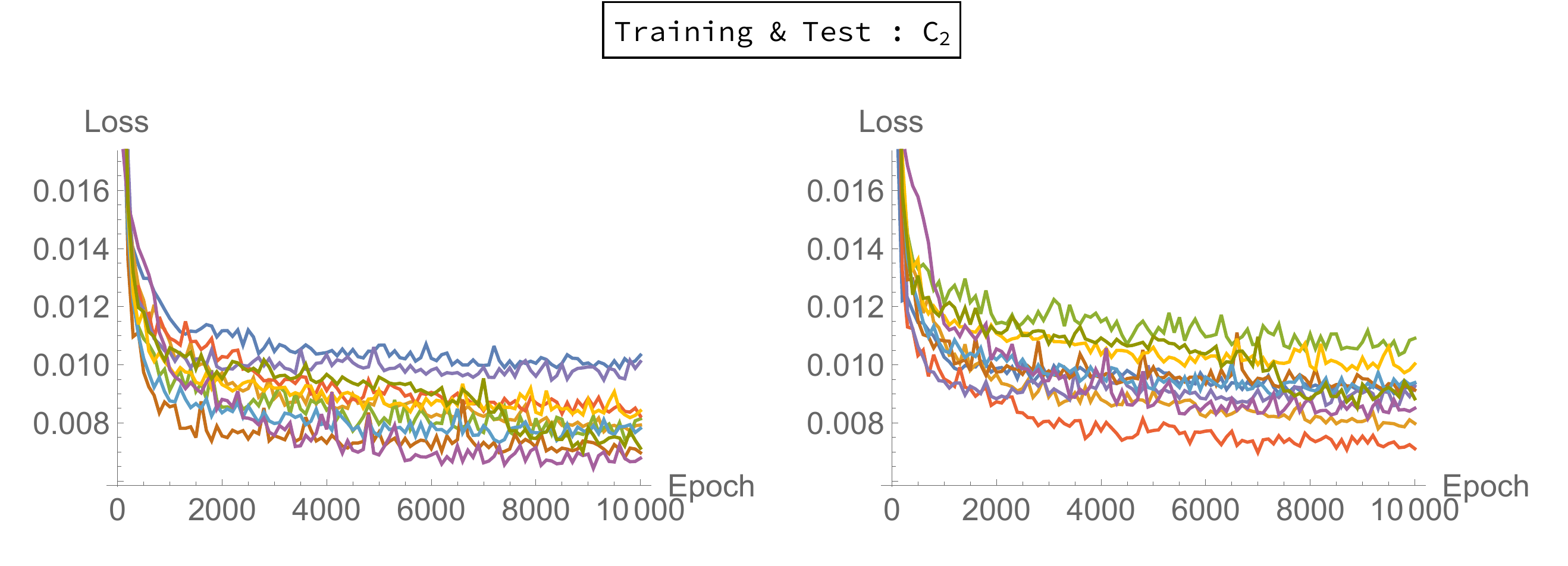}
}
\centerline{
\includegraphics[width=\columnwidth]{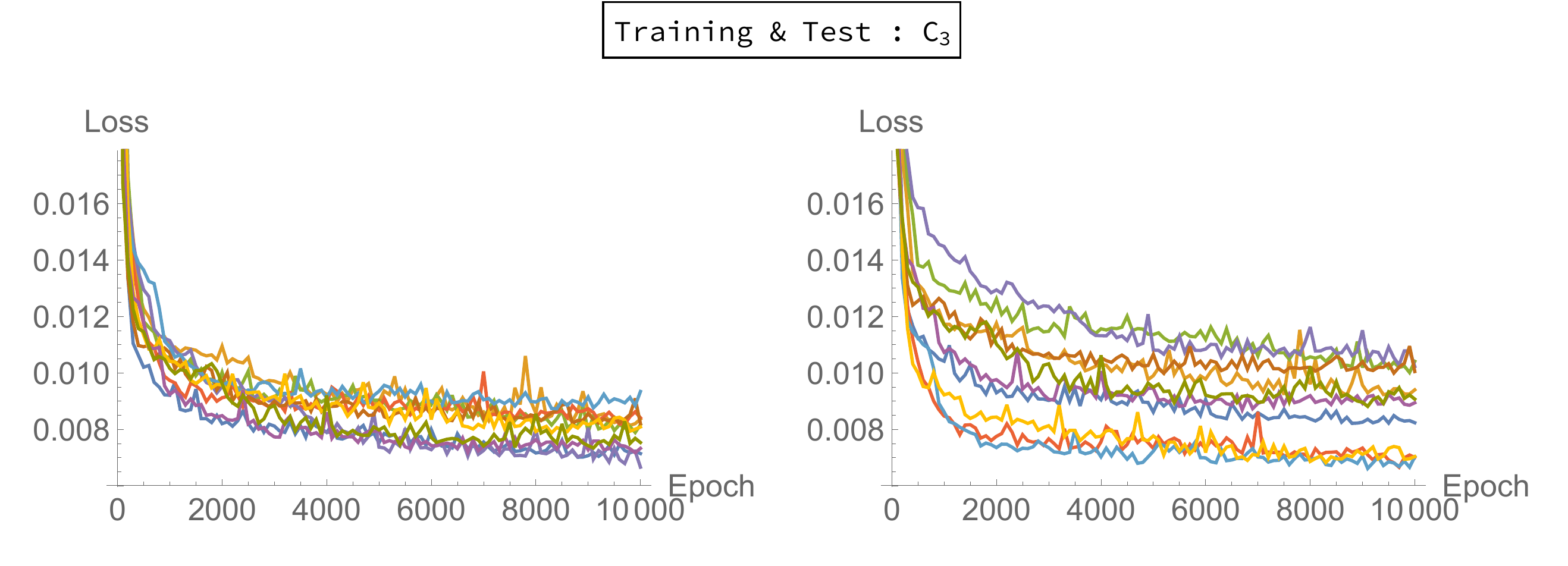}
}
\caption {\small
Evolution of the loss function.
The different panels correspond to $C_1$, $C_2$, and $C_3$ 
used as the training $\&$ test dataset from top to bottom.
The left panels show the loss function for the training dataset,
while the right panels show for the test dataset.
Different colors show the $10$ machines mentioned in Sec.~\ref{subsec:Setup}.
This figure corresponds to the setup in Fig.~\ref{fig:S__}.
}
\label{fig:Loss__}
\vskip 0.1in
\centerline{
\includegraphics[width=\columnwidth]{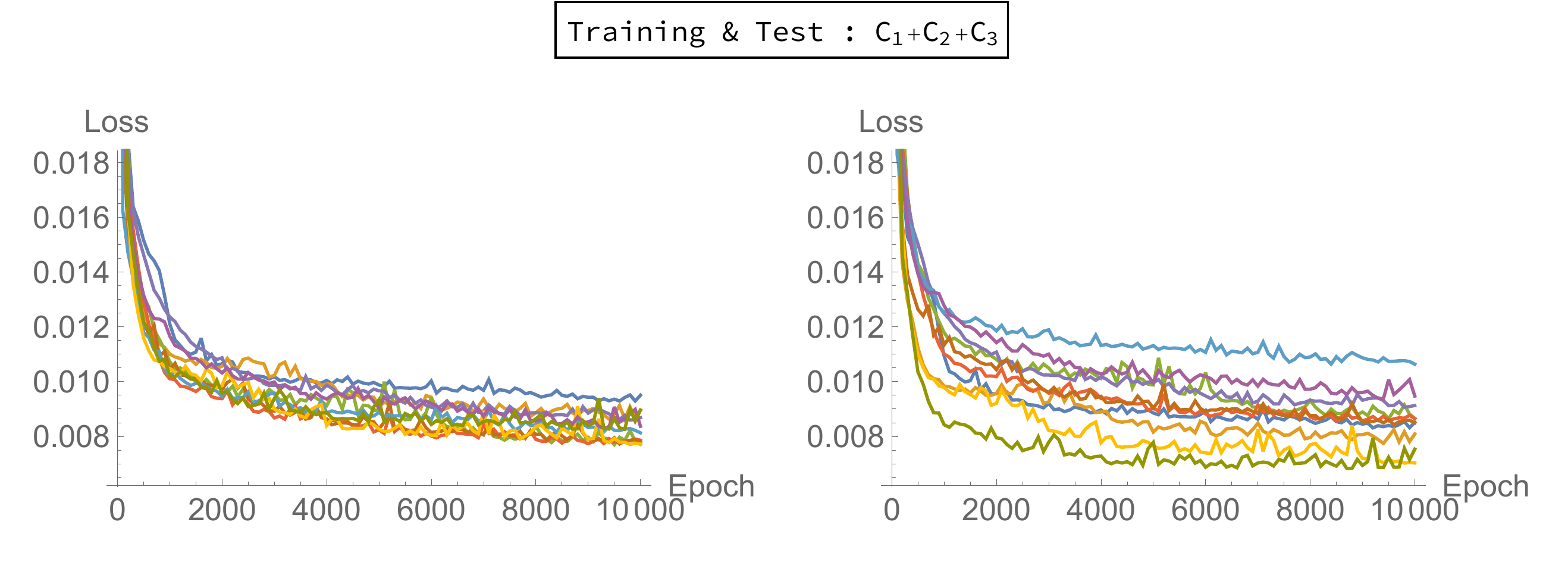}
}
\caption {\small
Same as Fig.~\ref{fig:Loss__}, except that $C_1 + C_2 + C_3$ is used for the training $\&$ test dataset.
This figure corresponds to the setup in Fig.~\ref{fig:S______}.
}
\label{fig:Loss______}
\end{figure}

In this appendix we show the evolution of the loss function as a function of the epoch elapsed.
Fig.~\ref{fig:Loss__} is the evolution
for the case with $C_1$, $C_2$, and $C_3$ used for the training $\&$ test dataset from top to bottom.
This figure corresponds to the setup of Fig.~\ref{fig:S__}.
It is seen that the loss function more or less converges at epoch $10,000$,
and also that there is no significant overfitting.

\begin{figure}
\centerline{
\includegraphics[width=\columnwidth]{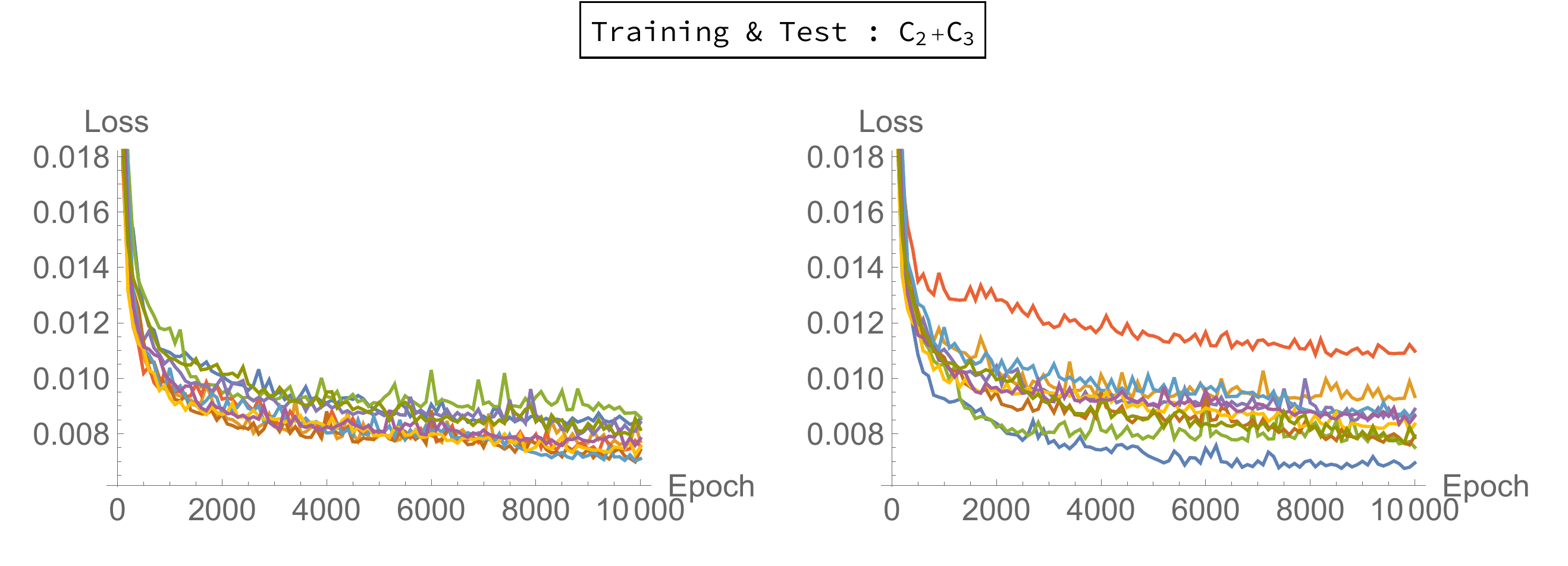}
}
\centerline{
\includegraphics[width=\columnwidth]{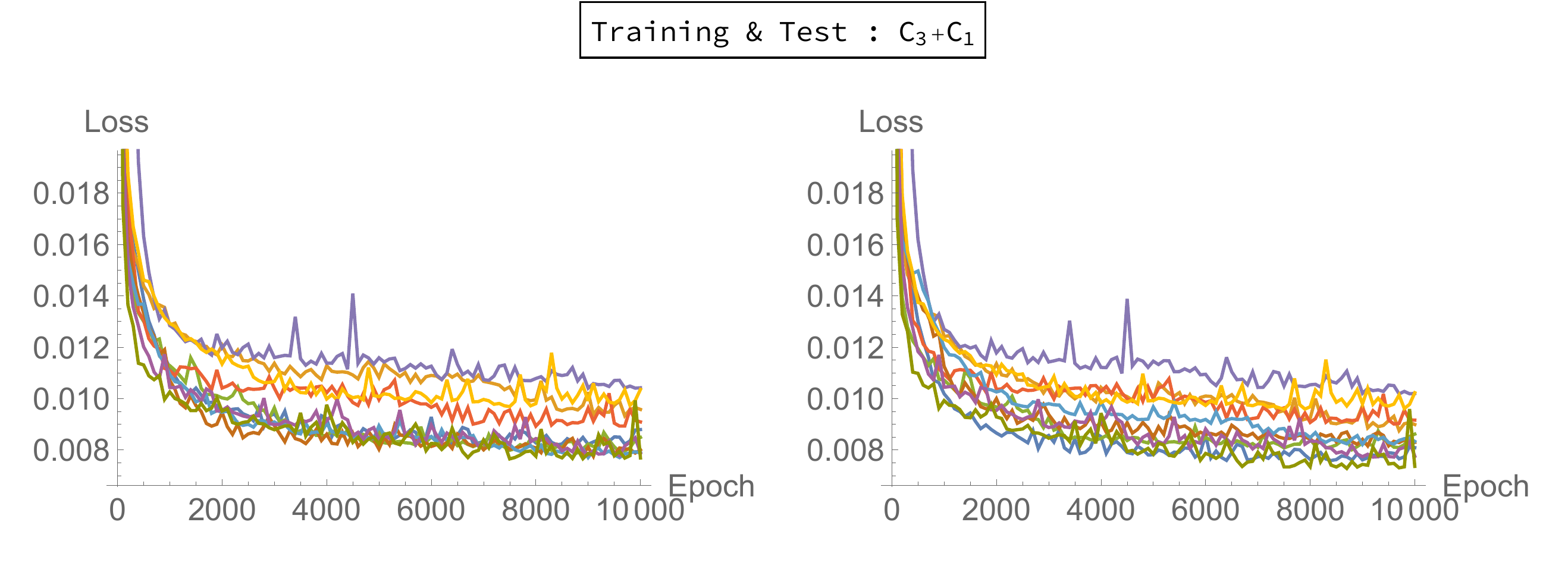}
}
\centerline{
\includegraphics[width=\columnwidth]{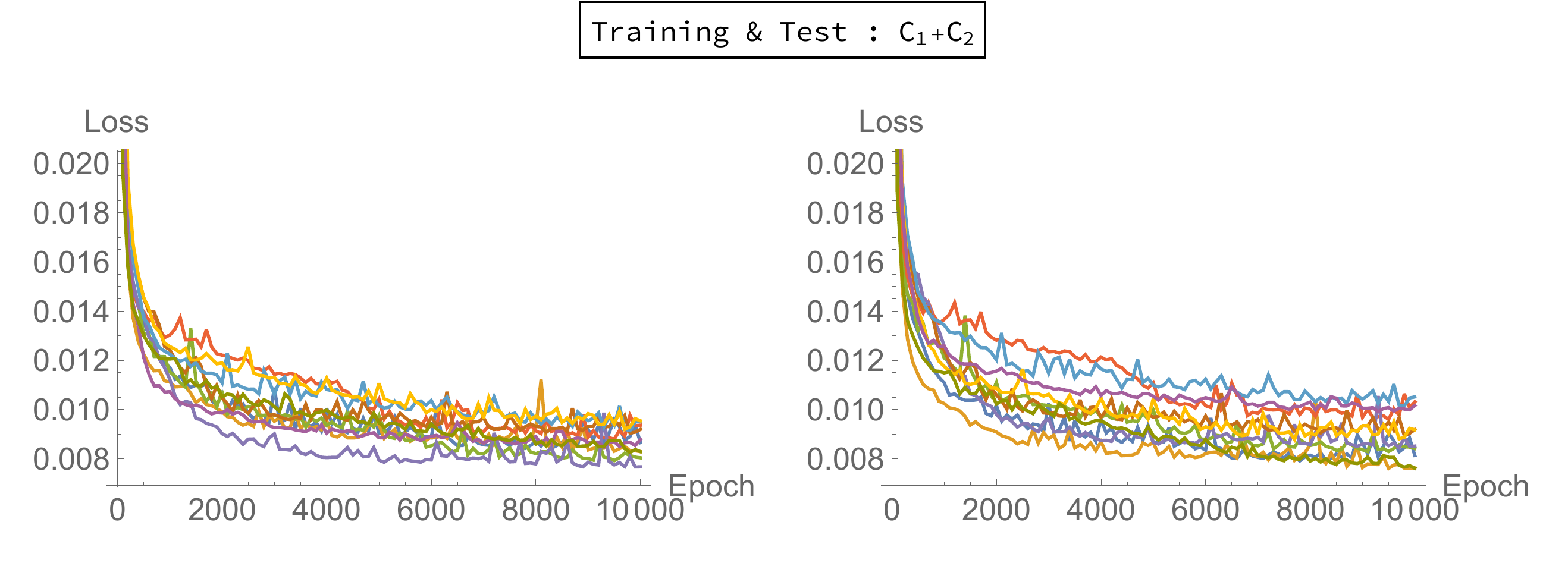}
}
\caption {\small
Same as Fig.~\ref{fig:Loss__}, except that $C_2 + C_3$, $C_3 + C_1$, and $C_1 + C_2$ are used 
for the training $\&$ test dataset.
This figure corresponds to the setup in Fig.~\ref{fig:S___}.
}
\label{fig:Loss___}
\end{figure}

We also show the evolution of the loss function 
for the cases corresponding to Figs.~\ref{fig:S______} and \ref{fig:S___}
in Figs.~\ref{fig:Loss______} and \ref{fig:Loss___}, respectively.
It is again seen that the loss function converges and also that there is no significant overfitting.

\clearpage

\clearpage

\small
\bibliography{ref}

\end{document}